\documentclass[aps,pra,shortbibliography,twocolumn,showpacs,superscriptaddress]{revtex4-1}

\usepackage{appendix}
\usepackage{amsmath}
\usepackage{graphicx}
\usepackage{amssymb}   %
\usepackage{dcolumn}
\usepackage{bm}
\usepackage{comment}
\usepackage{multirow}
\usepackage{array}
\usepackage{color}
\usepackage{physics}
\usepackage[colorlinks,citecolor=blue,linkcolor=blue,urlcolor=blue]{hyperref}
\usepackage{mathtools} 
\usepackage{braket}
\usepackage[normalem]{ulem} 
\usepackage{xcolor,cancel} 
\DeclareMathAlphabet \mathbfcal{OMS}{cmsy}{b}{n}

\graphicspath{ {Figures/} }

\begin{document}

\title{Photoinduced Nonperturbative Valley Polarization in Graphene }

\author{Hamed Koochaki Kelardeh}
\email{hkelardeh@pks.mpg.de}
\affiliation{Max Planck Institute for the Physics of Complex Systems, N\"othnitzer Straße 38, D-01187 Dresden, Germany}

\author{Mohammadreza Eidi}
\affiliation{Max Planck Institute for the Physics of Complex Systems, N\"othnitzer Straße 38, D-01187 Dresden, Germany}

\author{Takashi Oka}
\affiliation{The Institute for Solid State Physics, The University of Tokyo, Japan}

\author{Jan Michael Rost}
\affiliation{Max Planck Institute for the Physics of Complex Systems, N\"othnitzer Straße 38, D-01187 Dresden, Germany}

\date{\today}

\begin{abstract}
	
We investigate a valleytronic device based on graphene with charge separation at different sublattices and correspondingly at nonequivalent valleys. We characterize the maximality condition of valley polarization and investigate the parameters and conditions upon which we can coherently control the carriers and store data via valley degree of freedom. The valley polarization is controlled by the amplitude as well as the carrier-envelope phase of the pulse – one cycle optical field - and the curvature of the electron trajectory in the reciprocal space. 
When strong-field excitation is taken into account, the optical selection rule in perturbative optics is replaced by the nonadiabatic geometric effects. As a result, a nonperturbative valley polarization in two dimensional Dirac materials is induced regardless of having an intrinsic bandgap.
Microscopically, such a nonreciprocal response of graphene in the chiral electric field is encoded by the quantum Berry phase, as a (pseudo) magnetoelectric monopole.\\


\end{abstract}


\maketitle

\section{\label{Sec: Introduction} Introduction}

Today, geometric effects and controlling various phases of matter from superconducting to semiconducting states to topological insulators with conducting edge or surface states is a central topic in condensed matter physics with potential applications in quantum computing and room temperature superconductivity \cite{Xiao2010,Qi2011}. 

Among different quantum states, valley degree of freedom and manifestation of nontrivial quantum phases together with the generation of topological states of light has lately drawn significant attentions \cite{Schaibley2016,Kelardeh2016_Attosecond,Khanikaev2017,Vitale2018,Galan2019,Hafezi2019}.
The possibility of incorporating valley degree of freedom to store and carry information has drawn numerous attention and led to substantial electronic applications labeled as valleytronics. The valley pseudospin has potential to serve as a robust quantum signal processor and data storage with a petahertz bandwidth.

Theoretical and experimental investigation of Valley polarization have predominantly been  performed on insulating state such as monolayer hexagonal Boron Nitride \cite{Song2017}, and Diamond \cite{Isberg2013}, as well as semiconductors such as transition metal dichalcogenides (TMDC’s) \cite{Cao2012,Mak2012,Jones2013,Yoshikawa2019,Zeng2012,Heinz2017,Langer2018}, and Silicon \cite{Salfi2014}. Besides, the semimetallic systems with broken inversion symmetry such as  Bismuth \cite{Zhu2012,Zhu_Behnia2017}, gapped graphene \cite{Motlagh2019_gapped}, bilayer graphene \cite{Shimazaki2015,Sui2015,Kumar2020}, and graphene superlattice \cite{Yankowitz2012,Gorbachev2014} have been suggested as potential candidates for valley-contrasting optoelectronic devices. 
It has been frequently reported that the presence of a gap is a precondition for having valley polarization \cite{Yao2008,Lensky2015,Motlagh2019_gapped}. In this report, we are going to reexamine such a widely accepted notion and address the following important question: can we expect a sizable electric field-induced VP in gapless Dirac semimetals? 
Answering this question is imperative considering the fact that introducing an eV-scale staggered  bandgap ( also called Semenoff mass \cite{Semenoff1984}) in graphene, which is required for having a measurable VP in noncentrosymmetric graphene systems is technologically challenging. 
 
We report VP in gapless graphene with efficiency as high as 35$\%$ by utilizing a chiral optical pulse with few-cycle (sub ten femtosecond) and strong-field (volt per angstrom scale)  with controlled polarity, amplitude, and carrier-envelope phase (CEP). Such a noticeable VP in monolayer graphene is attributed to the nonperturbative dynamics of electrons in the strong electric field of the laser pulse, global topology of the honeycomb band structure and the nonequivalent Berry phase of $\pm\pi$ at the K and K$'$ valleys. 

In a nonperturbative nonlinear light-matter interaction where the light pulse has a strong amplitude (comparable with the interatomic field) and contains only one or two cycles, the incident electric field breaks the conservation law and gives rise to nonadiabatic topological effects and generation of the valley-contrasting population in the reciprocal space.

In a honeycomb crystal, the two degenerated valleys can be distinguished by pseudovector quantities, namely the Berry phase, connection, and curvature. Berry phase 
describes the phase of quantum mechanical wave functions accumulated along a closed loop in the reciprocal space (or any parametric space). Such a topological phase defines the topology of electronic states and plays a fundamental role in many emerging phenomena in condensed-matter systems, such as High-order harmonic generation \cite{Silva2019,Chacon2020,Gaarde_PRL2020,Bauer2020}, quantum spin Hall effect \cite{Herrero2018}, anomalous quantum Hall effect \cite{Nagaosa2010,Rubio2019,Cavalleri2020}.
Moreover, Berry connection which is defined as $ {\mathbfcal A}^{{\rm{nn}}} = \left\langle {{\phi _n}|i\grad |{\phi _n}} \right\rangle $  in the reciprocal space with $\grad  = \left( {{\partial _{{k_x}}},{\partial _{{k_y}}}} \right)$, and  ${{\phi _n}}$  being the periodic part of the Bloch band $n$, acts as the vector potential in the momentum space. $ {{\mathbfcal A}^{\rm{n}}}$ is related to the overlap of the two Bloch wavefunctions neighboring in the momentum space, and has the geometrical meaning of “connection” of the manifold in Hilbert space. The real space picture of Berry connection, in two-band $n,m$ with the presence of the strong field, is shifting the electron wave packets in the unit cell by a vector equal to ${{\mathbfcal A}^{n}}-{{\mathbfcal A}^{m}}$.

We discern that our proposed method using a single-cycle pulse for optical control and manipulation of the Valleys in pristine graphene is favored over the two-color bicircular field being pursued in another group \cite{Galan2019}.
The latter shaped-pulse lasers press for multiples of field cycles; thereby the excitonic and many-electron effects may hinder its practical application in valleytronics. Indeed, recent experiments bear witness to the formation of ultrafast exciton on sub-ps time scales in two-dimensional materials \cite{Steinleitner2017,Bernhard2018}.

Furthermore, we note that in order to gain VP close to one, the system requires to have an intrinsic energy gap at the band edge closely in-resonant with the pulse energy. In these systems, such as TMDC's, the valley pseudospin is associated with nonzero Berry curvature and an intrinsic magnetic moment near the Dirac cones. In other words, such crystals allow the simultaneous breakage of time-reversal symmetry ( by taking circular pulse) an inversion symmetry (by an intrinsic gap). Nevertheless, realizing a substantial VP predicted in this report is important, considering the abundance and fabrication challenges of TMDC nanostructures. We trust that such a noticeable VP in graphene can be essentially measured by state of the art technology \cite{Wang2013,Higuchi2017}.

While the physical mechanisms of the phenomena and effects will be discussed below in detail, we emphasize here that the nonperturbative nonreciprocal responses joint together some of the most fundamental issues in condensed matter physics, such as symmetries, topological nature of electrons, and electron correlation.

\section{Methodology}
\label{sec:method}

In this section, we calculate the electron dynamics in gapless graphene induced by a time-dependent, spatially-uniform electric field.
The microscopic theory of strong-field dynamics in solids, and the Coulomb-induced many body interactions is fundamentally described by the density matrix equations. 

We consider the field-matter interaction in the length gauge where the Houston functions (also known as the accelerated Bloch states) \cite{Houston_1940} have been previously utilized to describe the coherent dynamics of the various classes of crystalline materials by solving the time dependent density matrix equation. The benefit of using such basis sets is that we can obtain a separation of the induced current into intra- and inter-band components.

It is known that the photoexcited carriers scatter extremely fast in graphene compared to other materials, due to its linear energy dispersion \cite{Gierz2013_Snapshots}.
Li et al. \cite{Li2012} experimentally demonstrated that an ultrafast population inversion and broadband optical gain is established within the duration of a 35-fs light pulse.
Since time dependent Schrödinger equation (TDSE) does not incorporate the relaxation processes, in this study we derive the density matrix equations applicable for a description of dephasing and dessipation of electrons in the two-band tight-binding configuration of graphene. 

Here we only take the electron-electron interaction into consideration for our open quantum system, since other interactions such as phonon coupling or quantum optical effects possess time scales with orders of magnitude slower than the electronic dephasing and decoherence time scale \cite{Huttner2017}.\\

Let us consider the solution to the Schrödinger equation for a two-band crystal
in the presence of an intense optical field. The state of the system is described in terms of the general wave function $\Psi $, which obeys the Schrödinger equation
\begin{equation}
\label{Eq:TDSE}
{i\hbar \frac{{d\Psi }}{{dt}} = {\cal H}\Psi }
\end{equation}

with the Hamiltonian operator ${\cal H}$ given by

\begin{equation}
\label{Eq:Hamiltonian}
{{\cal H} = {{\cal H}_0} + e{\bf{F}}(t) \cdot {\bf{r}}}
\end{equation}

where $\mathbf F(t)$ is the  pulse's electric field, $e$ is the electron charge, and ${\cal H}_0$ is the Hamiltonian of the solid in the absence of the optical field. 
The position operator $\bf{r}$ in the reciprocal space representation is ${\bf{r}} = i{\grad _{\bf{q}}}$. 
We approximate ${\cal H}_0$ as a nearest-neighbor tight-binding (TB) Hamiltonian,
\begin{equation}
\label{G_Hamiltonian}
{\cal H}_0 = \left( {\begin{array}{*{20}{c}}
	\delta &{g({\bf{q}})}\\
	{{g^*}({\bf{q}})}&{ - \delta }
	\end{array}} \right)
\end{equation}

$\delta$ determines the slight doping energy of graphene ($ \sim $ meV), $g({\bf{q}}) = \gamma f({\bf{q}})$, with hopping integral $\gamma=-3.03$ eV and

\begin{equation}
f({\bf{q}}) = \exp \left( {i\frac{{a{q_x}}}{{\sqrt 3 }}} \right) + 2\exp \left( { - i\frac{{a{q_x}}}{{2\sqrt 3 }}} \right)\cos \left( {\frac{{a{q_y}}}{2}} \right)
\end{equation}

where $a=2.46~\mathrm{\AA}$ is the lattice constant. We define $g({\bf{q}}) = \left| {g({\bf{q}})} \right|{e^{i{\phi _{\bf{q}}}}}$, with ${\phi _{\bf{q}}} = {\tan ^{ - 1}}\left( {\frac{{{\mathop{\rm Im}\nolimits} (g({\bf{q}}))}}{{{\mathop{\rm Re}\nolimits} (g({\bf{q}}))}}} \right)$.

Accordingly, the eigenstates and eigenenergies of the conduction and valence bands can be found from the above Hamiltonian, ${\cal H}_0$, as follows 

\begin{equation}
\label{Eq:eigenstate}
\phi _{\bf{q}}^{c/v}({\bf{r}}) = \frac{{e^{i{\bf{q}} \cdot {\bf{r}}}}}{{\sqrt {2{E_{c/v}}(\delta + {E_{c/v}})} }}\left( {\begin{array}{*{20}{c}}
	{\delta + {E_{c/v}}}\\
	{\left| g \right|{e^{ - i{\phi _{\bf{q}}}}}}
	\end{array}} \right)
\end{equation}

\begin{equation}
\label{eigenenergy}
{E_{c/v}}({\bf{q}}) =  \pm \sqrt {{\delta ^2} + {{\left| {g({\bf{q}})} \right|}^2}} 
\end{equation}

where the +/- signs correspond to the conduction (c)/ valence (v) band, respectively.

An applied electric field generates both the intraband (adiabatic) and interband (nonadiabatic) electron dynamics. The intraband dynamics is determined by the Bloch acceleration theorem in the reciprocal space, $\hbar {\bf{\dot k}} = e{\bf{F}}({\bf{t}})$.  
For an electron with initial momentum $\bf{q}$ the electron dynamics is described by the time dependent wave vector given by ${\bf{k}}(t) = {\bf{q}} - e{\hbar ^{ - 1}}{\bf{A}}(t)$ with  ${\bf{A}}(t) =  - \int_{ - \infty }^t {\bf{F}} (t')dt'$ as the vector potential of the laser field. In fact, the electron wave packet with initial crystal wave vector $\bf{q}$ transforms to a trajectory-guided vector,${\bf{q}} \mapsto {\bf{k}}(t)$, where its  time-dependent trajectory is governed by the vector potential of the incident pulse. 

The most general solution to the time-dependent Schrödinger equation (\ref{Eq:TDSE}) can be written in the interaction representation as 

\begin{equation}
\label{Eq:Psi_expand}
{\Psi _{\bf{q}}}({\bf{r}},t) = \sum\limits_{n = v,c} {\int_{BZ} {\beta _{_{{\bf{k}}(t)}}^n\Phi _{n,{\bf{q}}}^{({\rm{H}})}({\bf{r}},t)} }
\end{equation}

where $\Phi _{n,{\bf{q}}}^{({\rm{H}})}({\bf{r}},t) = \phi _{{\bf{k}}({\bf{t}})}^n({\bf{r}}){e^{ - \frac{i}{\hbar }\int_{ - \infty }^t {E_n^T[{\bf{k}}(t')]dt'} }}$ are the time-dependent adiabatic basis set, i.e., the Houston functions and are the solutions of Schrödinger equation within a single band without an interband coupling. 

\begin{equation}
\label{Eq:E_T}
E_n^T[{\bf{k}}(t')] = {E_{\rm{n}}}[{\bf{k}}(t')] + e{\bf{F}}(t') \cdot{\mathbfcal{A}^{({\rm{nn}})}}[{\bf{k}}(t')]
\end{equation}

is called the modified band energy, accounting the dynamic phase [first term], as well as the geometric (Berry) phase [second term]. Here  $ {{\mathbfcal A}^{(nn)}} = i\left\langle {\phi _{{\bf{k}}({\bf{t}})}^n|{\grad_{\bf{q}}}|\phi _{{\bf{k}}({\bf{t}})}^n} \right\rangle $ is the global energy of ${n^{\rm{th}}}$-band including the Bloch energy dispersion [First term], as well as the field-induced geometrical counterpart [second term].
The later term is critical to detect the topological information of solid state systems including the peculiarities observed in the quantum Hall effect regime and pseudospin-related Berry’s phase \cite{Kelardeh2017_superlattice}.
  
$\phi _{{\bf{k}}({\bf{t}})}^n({\bf{r}})$ are the Bloch eigenstates of graphene (Eq. \ref{Eq:eigenstate}), and $n=v,c$ stands for the valence and conduction bands, respectively. 
The expansion coefficients ${\beta _c}(t)$ and ${\beta _v}(t)$  in Eq. \ref{Eq:Psi_expand} interpret as the probability amplitudes that at time $t$ the electron (or hole) is in the conduction or valence band. They satisfy the following system of integro-differential equations
\begin{equation}
\begin{array}{*{20}{l}}
{i\hbar \dot \beta _{{\bf{k}}(t)}^c = e{\bf{F}}(t) \cdot {\bf{Q}}_{{\bf{k}}(t)}^{cv}\beta _{{\bf{k}}(t)}^v}\\
{i\hbar \dot \beta _{{\bf{k}}(t)}^v = e{\bf{F}}(t) \cdot {\bf{Q}}_{{\bf{k}}(t)}^{cv*}\beta _{{\bf{k}}(t)}^c}
\end{array},
\end{equation}

where

\begin{equation}
\label{Eq:Q_cv}
{\bf{Q}}_{{\bf{k}}(t)}^{cv} = {\mathbfcal A}_{{\bf{k}}(t)}^{cv}{e^{\frac{i}{\hbar }\int_{ - \infty }^t {\left( {E_c^T[{\bf{k}}(t')] - E_v^T[{\bf{k}}(t')]} \right)dt'} }}
\end{equation}

determines the  matrix  element  of  interband interaction where ${\mathbfcal A}_{{\bf{k}}(t)}^{cv}$ is the interband Berry connection. The interband Berry connection can be defined in terms of the transition dipole moments (TDM) ${\bf{D}}_{{\bf{k}}(t)}^{cv}$ \cite{Kelardeh2015_Graphene} as ${\bf{D}}_{{\bf{k}}(t)}^{cv} = e{\mathbfcal A}_{{\bf{k}}(t)}^{cv}$. TDM determines optical transitions between the VB and CB at a crystal momentum $\mathbf q$. 
The exponential factor in Eq. \ref{Eq:Q_cv} which is calculated from Eq. \ref{Eq:E_T}, defines the global  phase difference between  the  states  of  the CB and VB (i.e., the generalized band offset).

We can write the intraband and interband components of the Berry connection operator In the two-band graphene system, the Berry connection has the following matrix form

\begin{equation}
\mathbfcal{\hat A} (\mathbf{q})= \left( {\begin{array}{*{20}{c}}
	{{\mathbfcal{A}^{cc}}}&{{\mathbfcal{A}^{cv}}}\\
	{\mathbfcal{A}^{cv*}}&{{\mathbfcal{A}^{vv}}}
	\end{array}} \right)
\end{equation}

with 

\begin{equation}
\label{Eq:A_nm}
{{\mathbfcal A}^{nm}} = i\left\langle {\phi _{\bf{q}}^{(n)}|{\grad }|\phi _{\bf{q}}^{(m)}} \right\rangle 
\end{equation}

Substituting Eq. \ref{Eq:eigenstate} into Eq. \ref{Eq:A_nm}, one finds the an analytical expression for the matrix elements of the Berry connection in the tight-binding approximation. 

The intraband components read

\begin{eqnarray}
\label{Eq:BerryCon_intra}
\begin{array}{l}
A_x^{cc/vv} = \frac{{a{\gamma ^2}}}{{\sqrt 3 }}\frac{{1 + {c_0}({c_3} - 2{c_0})}}{{{u_{c/v}}}}\\
A_y^{cc/vv} = a{\gamma ^2}\frac{{{s_0}{s_3}}}{{{u_{c/v}}}}
\end{array}
\end{eqnarray}

and interband Berry connection:

\begin{eqnarray}
\label{Eq:BerryCon_interband}
\begin{array}{l}
{\cal{A}}_x^{cv} = \frac{{a{\gamma ^2}}}{{2\sqrt 3 {E_c}\left| g \right|}}\left[ {1 + {c_0}({c_3} - 2{c_0})} \right] + i\frac{{\sqrt 3 \delta a{\gamma ^2}}}{{2E_c^2\left| g \right|}}{c_0}{s_3}\\
{\cal{A}}_y^{cv} = \frac{{a{\gamma ^2}}}{{2{E_c}\left| g \right|}}{s_0}{s_3} + i\frac{{\delta a{\gamma ^2}}}{{2E_c^2\left| g \right|}}{s_0}\left( {{c_3} + 2{c_0}} \right)
\end{array}
\end{eqnarray}

where ${u_{c/v}} = \left| {{g^2}} \right| + {(\delta + {E_{c/v}})^2}$, and ${c_0} = \cos \left( {a{k_y}/2} \right)$, ${s_0} = \sin \left( {a{k_y}/2} \right)$, ${c_3} = \cos \left( {\sqrt 3 a{k_x}/2} \right)$,  ${s_3} = \sin \left( {\sqrt 3 a{k_x}/2} \right)$. 

So far we treated the graphene system as a closed two-level system; now, we extend our model to incorporate dephasing and decoherence into the driven solid-state electron dynamics. To this end, we employ the Liouville von Neumann equation and propagate the reduced density matrix in the length gauge interaction picture to describe quantum dynamics in the presence of the relaxation process as below:

\begin{equation}
\label{Eq:Rho_Dot}
\left\{ {\begin{array}{*{20}{l}}
	{\dot \rho _{{\bf{k}}(t)}^{cv} = \frac{i}{\hbar }{\bf{F}}(t) \cdot {\bf{Q}}_{{\bf{k}}(t)}^{cv}\left[ {\rho _{{\bf{k}}(t)}^{cc} - \rho _{{\bf{k}}(t)}^{vv}} \right] - \gamma \rho _{{\bf{k}}(t)}^{cv}}\\
	{\dot \rho _{{\bf{k}}(t)}^{cc} = 2{\rm{Re}}\left[ {\frac{i}{\hbar }{\bf{F}}(t) \cdot {\bf{Q}}_{{\bf{k}}(t)}^{cv*}\rho _{{\bf{k}}(t)}^{cv}} \right] - \gamma \rho _{{\bf{k}}(t)}^{cc}}
	\end{array}} \right.
\end{equation}

${\rho^{ij}}$ are the solution of the Born-Markov Master equation by averaging over the thermal bath degrees of freedom:
\begin{equation}
\label{Eq:Rho}
\rho (t) = {\rm{T}}{{\rm{r}}_B}\left( {\left| \Psi  \right\rangle \left\langle \Psi  \right|} \right) = \left( {\begin{array}{*{20}{c}}
	{\rho _{{\bf{k}}(t)}^{cc}}&{\rho _{{\bf{k}}(t)}^{cv}}\\
	{\rho _{{\bf{k}}(t)}^{cv*}}&{\rho _{{\bf{k}}(t)}^{vv}}
	\end{array}} \right)
\end{equation}

The diagonal terms in Eq. \ref{Eq:Rho}  represent the band population $\rho _{{\bf{k}}(t)}^{cc} = {\left| {\beta _{{\bf{k}}(t)}^c} \right|^2}$ and the nondiagonal terms define the polarization function. Also note that ${\rho ^{vv}} = 1 - {\rho ^{cc}}$.

The relaxation rate $\gamma$ in Eq. \ref{Eq:Rho_Dot} has an inverse proportion of the dephasing time, ${\gamma _{({\rm{PHz}})}} = \frac{1}{{{T_{({\rm{fs}})}}}}$.  

The Liouville-Von Neumann equation (Eq.\ref{Eq:Rho_Dot}) in our settings goes beyond the Boltzman transport theory where only the band dispersion and the consequent group velocity appears in the equation.

We assume that the VB is fully occupied and the CB is empty. The applied pulse is an intense single optical oscillation. In the numerical simulation, the important precondition for the chosen electric field waveform impose that $\int_{ - \infty }^\infty  {{\bf{F}}(t')dt'}  = 0$. We characterize the laser waveform such that it satisfies this condition, by defining ${\bf{A}}(t)$ and obtaining ${\bf{F}}(t)$ as the temporal derivative of it. We employ the following vector potential waveform for the elliptically-polarized pulse.
\begin{equation}
\label{Eq:VectorPotential}
\begin{array}{l}
{A_x}(t) = {F_{0}}{\omega ^{ - 1}}{e^{ - {{(t/\tau )}^2}}}\sin (\omega t + {\phi _{\rm{CEP}}}),\\
{A_y}(t) = \pm {\varepsilon}{F_{0}}{\omega ^{ - 1}}{e^{ - {{(t/\tau )}^2}}}\cos (\omega t + {\phi _{\rm{CEP}}}),
\end{array}
\end{equation}

${F_{0}}$ is the field amplitude,  $\tau $ is the pulse length corresponding to carrier frequency $\omega$  = 1.5 eV/$\hbar $, and ${\phi _{\rm{CEP}}}$ is the CEP. Sign $ \pm $ determines the right and left-handedness of the pulse. $\varepsilon \in [0,1]$ controls the degree of ellipticity and hence the curvature of electron trajectory. \\

\begin{figure}   
	\centering   
	\includegraphics[width=.95\columnwidth,keepaspectratio ]{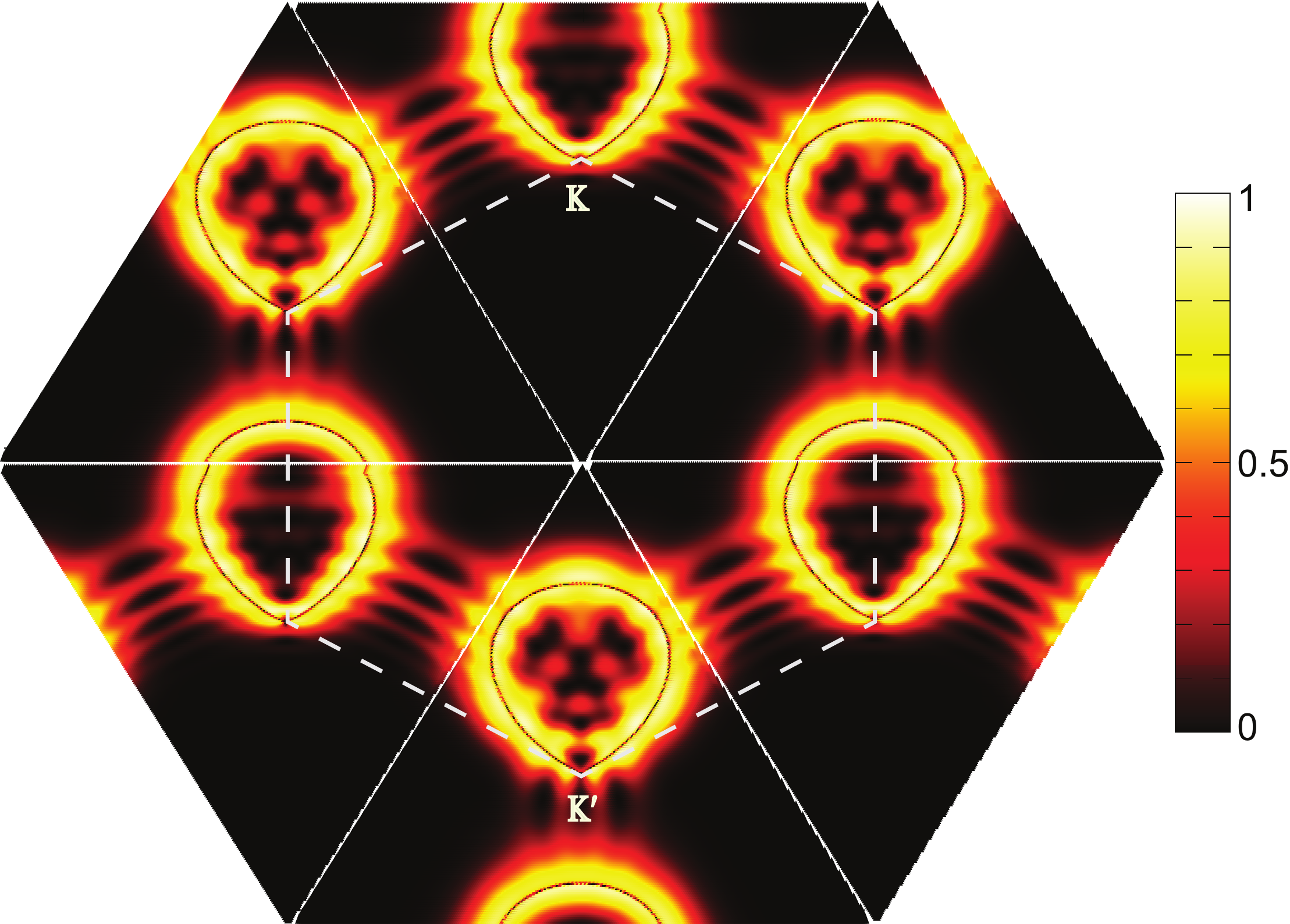}   \caption{Schematic illustration of the CB population excitation of graphene at the end of a single-cycle circular field with $F_0=0.9 {\rm{V/\AA}}$, CEP=0 and dephasing time T=100 fs. ${\rho _{{\bf{k}}(t \to {t_f})}^{cc}}$, which is calculated from Eq. \ref{Eq:Rho_Dot} is color mapped between 0 (no excitation states) to 1 (full occupation states) in the reciprocal space. The first BZ  and the inequivalent $\rm{K}$ and $\rm{K'}$ points are shown. The triangular borderline in-between the $\rm{K}$ and $\rm{K'}$ valleys are also indicated.  } 
	\label{Fig_ExtendedBZ_T_100_Hex} 
\end{figure}

System of equations \ref{Eq:Rho_Dot} determines the laser-induced electron dynamics;  solving these coupled diffrentio-integral equations, we obtain reciprocal space distribution of electrons in the conduction and valence bands. Correspondingly, the time-dependent CB population is defined by $\rho _{{\bf{k}}(t)}^{cc} = {\left| {\beta _{{\bf{k}}(t)}^c} \right|^2}$. The residual value of the CB population, ${\rho _{{\bf{k}}(t \to {t_f})}^{cc}}$, is defined as the population after the pulse. \\
We define the VP as

\begin{equation}
\label{Eq:VP}
{\rm{VP}} = \frac{{\left| {{n_{\rm{K}}} - n{}_{{\rm{K'}}}} \right|}}{{{n_{\rm{K}}} + n{}_{{\rm{K'}}}}},
\end{equation}
with

\begin{equation}
\label{Eq:NkNKprime}
\begin{array}{*{20}{l}}
{{n_{\rm{K}}} = \int_{ - {q_y}/\sqrt 3 }^{{q_y}/\sqrt 3 } {\int_0^{3\kappa /2} {\rho _{{\bf{k}}(t \to {t_f})}^{cc}{\rm{d}}{q_y}{\rm{d}}{q_x}} } ,}\\
{{n_{{\rm{K'}}}} = \int_{{q_y}/\sqrt 3 }^{ - {q_y}/\sqrt 3 } {\int_{ - 3\kappa /2}^0 {\rho _{{\bf{k}}(t \to {t_f})}^{cc}{\rm{d}}{q_y}{\rm{d}}{q_x}} } ,}
\end{array}
\end{equation}

as the total population of the K- and $\rm{K'}$ valleys inside their corresponding triangle and  $\kappa  = 4\pi /3{a_0}$ is a constant. Due to the honeycomb structure of graphene, the triangular meshgrid is chosen to fully take in the population proportions throughout the BZ, as illustrated in Fig. \ref{Fig_ExtendedBZ_T_100_Hex}.\\
The fine (eye-drop shaped) line within the bright colored CB momentum distribution represent the separatrix, which is the superposition of initial conditions where their trajectories pass through the Dirac point. In other words, the separatrix is the mirror symmetry of the vector potential (polarization state), that steer the trajectory of electron in the momentum space. \\

\section{Result and discussion}
\label{sec:result}

Valley degeneracy in the momentum space of the conduction and valence bands presents an additional degree of freedom for charge carrier manipulation. As its counterpart spin states in spintronics, controlling the population of valley states is essential to the development of valley-based electronics.

We show that the single-cycle pulse results in a significantly large VP. The controlling parameters  of the optical pulse are the field amplitude $F_0$, ellipticity (tuned by $\varepsilon$), and the carrier envelope phase ${\phi _{\rm{CEP}}}$. polarization state 


\subsection{\label{Sec:Field_amplitude} Field Amplitude and Ellipticity}

\begin{figure}   
\centering   
\includegraphics[width=.95\columnwidth,keepaspectratio ]{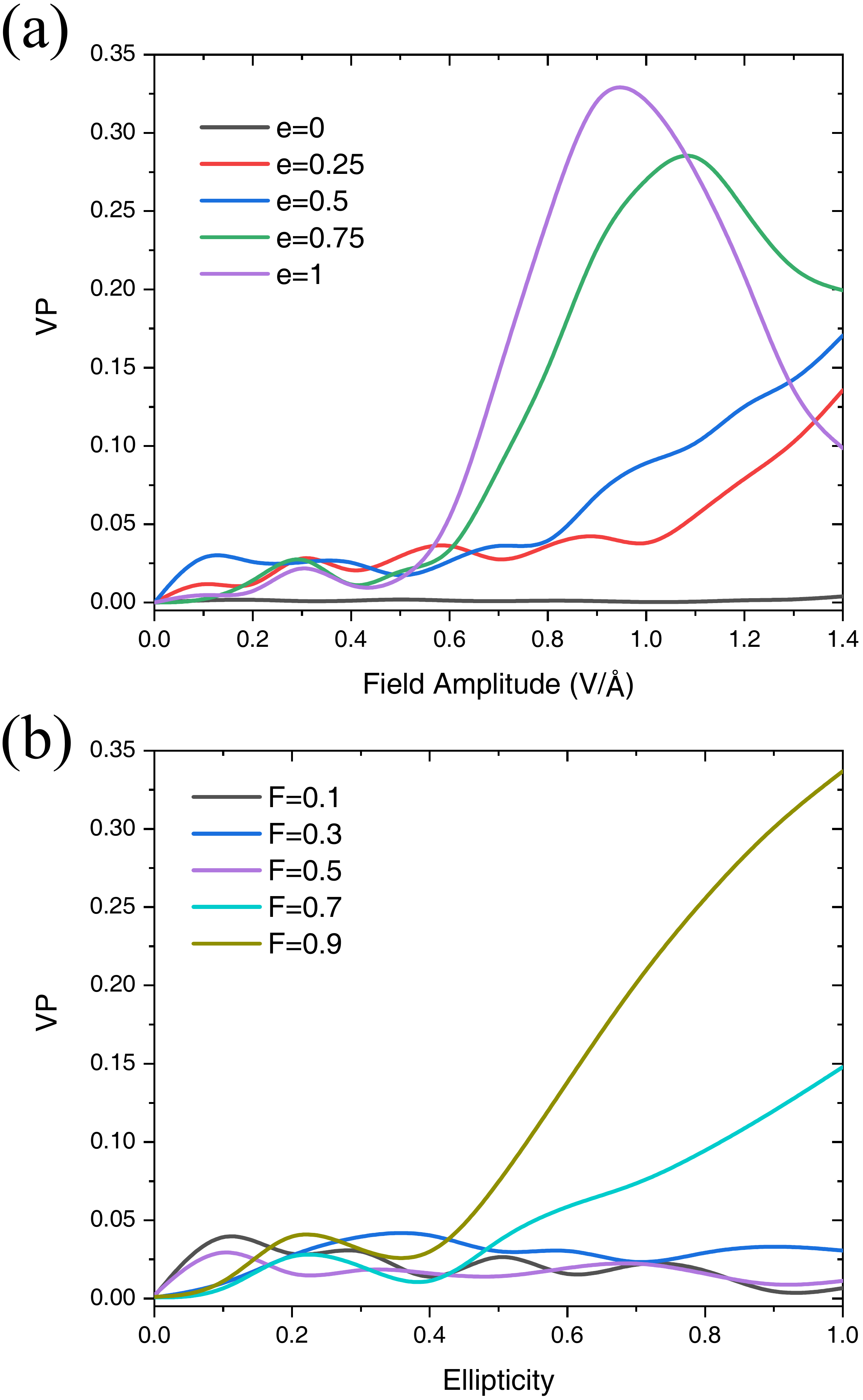}   \caption{(a) depicts the VP as a function of field amplitude, $F_0$, for different ellipticities  $\varepsilon$. the purple color corresponds to the circularly polarized pulse, in which the VP increases monotonically with field and reaches to a maximum of ~35 percent at ${F_0} = 0.9\,{\rm{V/\AA}}$, before sharply falling at higher fields. (b) VP is plotted versus ellipticity $\varepsilon$ for different field amplitude $F_0$. For linearly polarized pulse VP=0 since such pulse preserve the time-reversal symmetry. For $F_0 > 0.5 {\rm{V/\AA}}$, a threshold $\varepsilon \simeq 0.5 $ is observed above which the VP steeply increases. } 
\label{Fig_Combined_VP_vs_F_ECC} 
\end{figure}

Firstly, we look into the simultaneous roles of ellipticity and field amplitude on VP and find the optimal $\varepsilon$ and $F_0$ for which the VP is maximized. We set CEP=0 and exclude the relaxation processes in this section. Fig. \ref{Fig_Combined_VP_vs_F_ECC}(a) plots the VP versus $F_0$ for a range of polarization states from linear ($\varepsilon$=0) to circular pulse ($\varepsilon$=1). Evidently, since the linear pulse preserves the time-reversal symmetry, its corresponding VP is zero. 

We span through a wide range of $F_0$ up until 1.4 ${\rm{V/\AA}}$. In this field region, VP increases monotonically for $\varepsilon$=0.25 and 0.5, whereas for $\varepsilon$=0.75 and 1, VP after a slow and oscillatory increase at small fields, grows exponentially to its highest amount then falls at higher fields. The highest VP corresponds to the circular pulse ($\varepsilon=1$) with field amplitude ${F_0}  \simeq  0.9\,{\rm{V/\AA}}$. 

Fig. \ref{Fig_Combined_VP_vs_F_ECC}(b), on the other hand, plots the VP as a function of polarization state for different laser field amplitudes.  For $F_0 < 0.5 {\rm{V/\AA}}$, VP is approximately zero with practically no influence by the ellipticity modulation. However, at moderate fields $ \sim $ 0.5 to 0.9  ${\rm{V/\AA}}$, VP is abruptly enhanced for $\varepsilon  \ge 0.5$.


\subsection{\label{Sec:relaxation} Induced Current and Net Charge Transport} 

To understand the behavior of VP versus field amplitude and ellipticity, we calculate the generated photocurrent and net charge transfer through the graphene system. 
The time-dependent electric field of the optical pulse causes the polarization of the system which generates an electric current ${\bf{J}}(t) = \left\{ {{J_x}(t),{J_y}(t)} \right\}$.  Both intraband ($\mathbf{J}^{\rm{intra}}(t)$) and interband ($\mathbf{J}^{\rm{inter}}(t)$) currents contribute to the total current, $\mathbf{J}(t)=\mathbf{J}^{\rm{intra}}(t)+\mathbf{J}^{\rm{inter}}(t)$, in the system. the two-band density matrix formalism we incorporated in the main text results in the following equations for the interaband and interband currents:

\begin{equation}
\label{Eq:current}
\begin{array}{*{20}{l}}
{{{\bf{J}}_{{\mathop{\rm int}} ra}}(t) = 2e\sum\limits_{\alpha  = c,v} {\int_{\overline {{\rm{BZ}}} } {{{\bf{V}}^\alpha }} } \left( {{\bf{k}}({\bf{q}},t)} \right){\rho ^{\alpha \alpha }}({\bf{q}},t)d{\bf{q}},}\\
{{{\bf{J}}_{{\mathop{\rm int}} er}}(t) = 2e\int_{\overline {{\rm{BZ}}} } {{{\bf{V}}^{vc}}} \left( {{\bf{k}}({\bf{q}},t)} \right){\rho ^{cv}}({\bf{q}},t)d{\bf{q}} + {\rm{c}}.{\rm{c}}.}
\end{array}
\end{equation}

${{\rho ^{\alpha \alpha }}({\bf{q}},t)}$ is the $\alpha$\textsuperscript{th}-band occupation (with the index $m$ running over the valence and conduction bands, i.e.\ $\alpha=v$ and~$c$, respectively) and ${{\rho ^{cv}}({\bf{q}},t)}$ is the interband coherence. 
The factor of 2 in Eq. \ref{Eq:current} is due to the spin degeneracy. 

${{\bf{V}}^\alpha }$ and ${{\bf{V}}^{\alpha \alpha '}}$, ($\alpha$ and $\alpha'$ interchange between $c$ and $v$) are the matrix elements of the velocity operator where in the two-band picture has the following form:

\[{\bf{V}} = \left( {\begin{array}{*{20}{c}}
	{{{\bf{V}}^{cc}}}&{{{\bf{V}}^{cv}}}\\
	{{{\bf{V}}^{cv*}}}&{{{\bf{V}}^{vv}}}
	\end{array}} \right).\]

The intraband velocity, within the quantum kinetic theory, is defined as
 
\begin{equation}
\label{Eq:velocity_intra}
\begin{array}{l}
{{\bf{V}}^\alpha } = \frac{1}{\hbar }{\grad _{\bf{k}}}E_\alpha ^T[{\bf{k}}(t)] = \\
\,\,\,\,\,\,\,\,\,\,\frac{1}{\hbar }\left[ {{\grad _{\bf{k}}}{E_\alpha }[{\bf{k}}(t)] + e{\grad _{\bf{k}}}\left( {{\bf{F}}(t) \cdot {{\bf{A}}^{(\alpha \alpha )}}[{\bf{k}}(t)]} \right)} \right]
\end{array}
\end{equation}

and interband velocity matrix element as

\begin{equation}
\label{Eq:velocity_inter}
{{\bf{V}}^{\alpha \alpha '}}({\bf{k}}) = \frac{i}{\hbar }{\bf{Q}}_{{\bf{k}}({\bf{q}},t)}^{\alpha \alpha '}\left[ {E_\alpha ^T({\bf{k}}({\bf{q}},t)) - E_{\alpha '}^T({\bf{k}}({\bf{q}},t))} \right]
\end{equation}

${\bf{Q}}_{{\bf{k}}({\bf{q}},t)}^{\alpha \alpha '}$ in the above equation is obtained from Eq. \ref{Eq:Q_cv} and is related to the interband Berry connection.

It is important to note that the Berry connection and curvature effects come into play in the matrix elements of both inetra- and interband velocities. 

Intraband velocity contains two terms:  ${\bf{V}}_{\rm{gr}}^\alpha ({\bf{k}}) = {\grad _{{\bf{k}}{\kern 1pt} }}{E_\alpha }({\bf{k}})$ is the particle (i.e.\ electron or hole) group velocity with $E_\alpha({\bf k})$ the bands' energy dispersion, and
${\bf{V}}_{{\rm{anom}}}^\alpha ({\bf{k}}) = e{\grad _{\bf{k}}}\left( {{\bf{F}}(t) \cdot {{\bf{A}}^{(\alpha \alpha )}}[{\bf{k}}({\bf{q}},t)]} \right)$ is the anomalous velocity that captures the quantum geometry of the Bloch wavefunction and Berry phase \cite{Stephanov2012}.  ${\bf k}$ is the quasi momentum defined in terms of the crystal wave vector ${\bf q}$ and the vector potential ${\bf A}(t)$ of the laser's electric field as ${\bf{k}}({\bf{q}},t) = {\bf{q}} - e/\hbar {\bf{A}}(t)$.

The current induced by the single-cycle optical pulse, respectively, results in the charge transfer across the system, which can be calculated from the following expression

\begin{equation}
{\bf{Q}} = \int_{ - \infty }^\infty  {{\bf{J}}(t){\rm{d}}t} 
\end{equation}

\begin{figure}   
	\centering   
	\includegraphics[width=.95\columnwidth,keepaspectratio ]{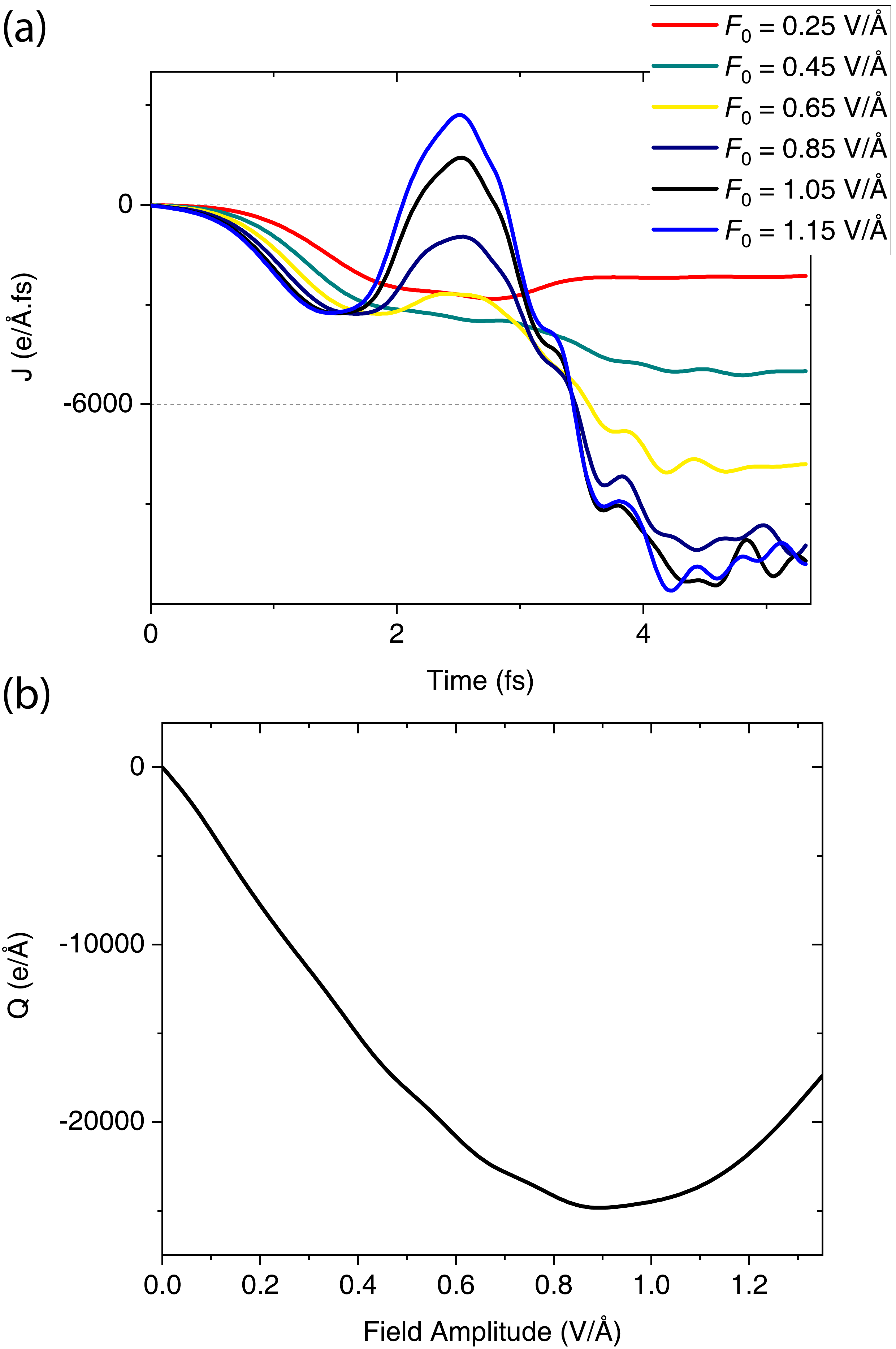}   \caption{ (a) photoinduced current in graphene is plotted within the incidence of the single-cycle circular field. The corresponding field amplitudes, $F_0$ associated with different colors are indicated. The current density oscillates at high field and changes sign at ${F_0} > 0.9 {\rm{V/\AA}}$ associated with the maximum valley polarization. (b)  Transferred charge density through graphene monolayer as a function of field amplitude $F_0$, also manifests the resulting field dependence of the valley-contrasting excitation. } 
	\label{Fig_J_Q} 
\end{figure}

Fig. \ref{Fig_J_Q}(a) plots the time-dependent current in graphene for the circularly polarized pulse with different Field amplitudes. 
From the numerical calculations, we observe that the current changes drastically, and this occurs in the same situation where the VP shows the switching.
The reason that the photocurrent and VP show the common behavior can be explained by the trajectory of e in the momentum space.
The dominant excitation taking place at $\rm{K}$-point will move as ${\bf{q}} - e/\hbar {\bf{A}}(t)$, and when they move to the $\rm{K'}$-valley, the group velocity, as well as the Anomalous velocity, will show the sign flip.
Such a sign change in the current density, occurs roughly at  $ F_0 \sim 0.9 {\rm{V/\AA}}$, where the maximum VP appears (c.f. Sec. \ref{Sec:Field_amplitude}). 

The residual population and current, in turn, translate to a transferred charge density and is plotted in Fig. \ref{Fig_J_Q}(b) as a function of the peak laser field. Respective to the sign change of current in Fig. (a), Q also reacts on the critical field and changes its slope at $ F_0 \sim 0.9 {\rm{V/\AA}}$. Such observable substantiate the VP field dependence as represented in Fig. \ref{Fig_Combined_VP_vs_F_ECC}(a).

\subsection{\label{Sec:relaxation} Relaxation Dynamics} 

In the previous section, we explored the variation of VP as a function of field amplitude and laser waveform, without taking the role of relaxation into account (i.e., $T  \to \infty $ in Eq. \ref{Eq:Rho_Dot}). In this and following sections, we study in what manner the relaxation mechanisms alter functionality and amplitude of the VP. 

\begin{figure}   
\centering   
\includegraphics[width=.95\columnwidth,keepaspectratio ]{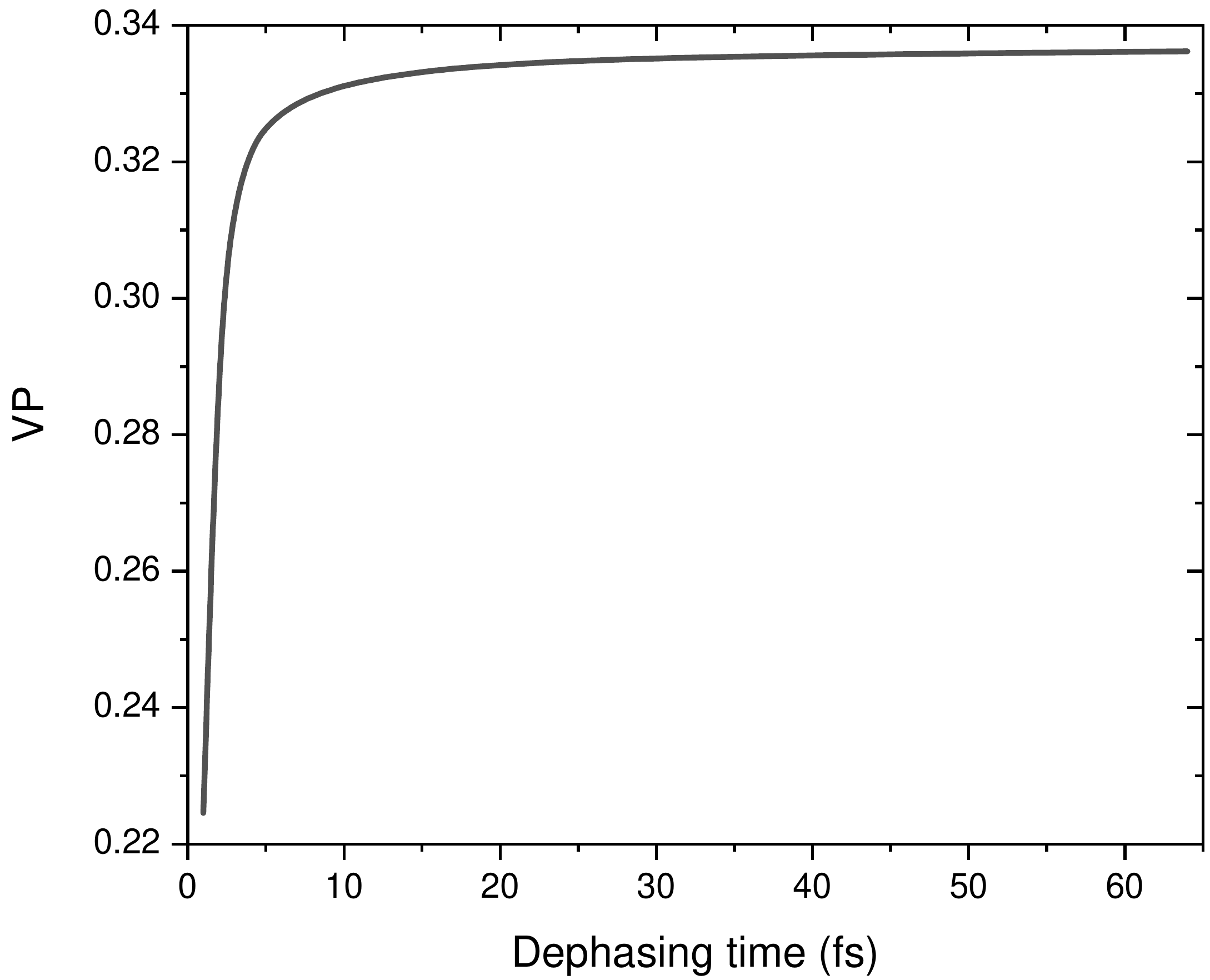}   \caption{ Valley polarization (VP) plotted as a function of the dephasing time. VP is practically unaffected by the relaxation up until $T \simeq 3$ fs where it exponentially decays to a magnitude of $\sim$0.22 at $T=1$ fs. The laser in this case is a single cycle circular pulse with ${F_0} = 0.9\,{\rm{V/\AA}}$ and CEP=0. } 
\label{Fig_VP_vs_T} 
\end{figure}

In previous section we showd that, in the absence of relaxation, the circular pulse with $F_0 = 0.9 {\rm{V/\AA}}$  gives rise to a maximum VP for a single-cycle optical laser. In Fig. \ref{Fig_VP_vs_T} we plot the VP as a function of dephasing time T, with field parameter corresponding to the maximum VP. As depicted, the VP falls suddenly for ultrafast dephasing time $T < 3$ fs. Notably, for the fastest relaxation time T=1 fs, we expect to have $~$22$\% $ of the VP.

\begin{figure}   
	\centering   
	\includegraphics[width=.95\columnwidth,keepaspectratio ]{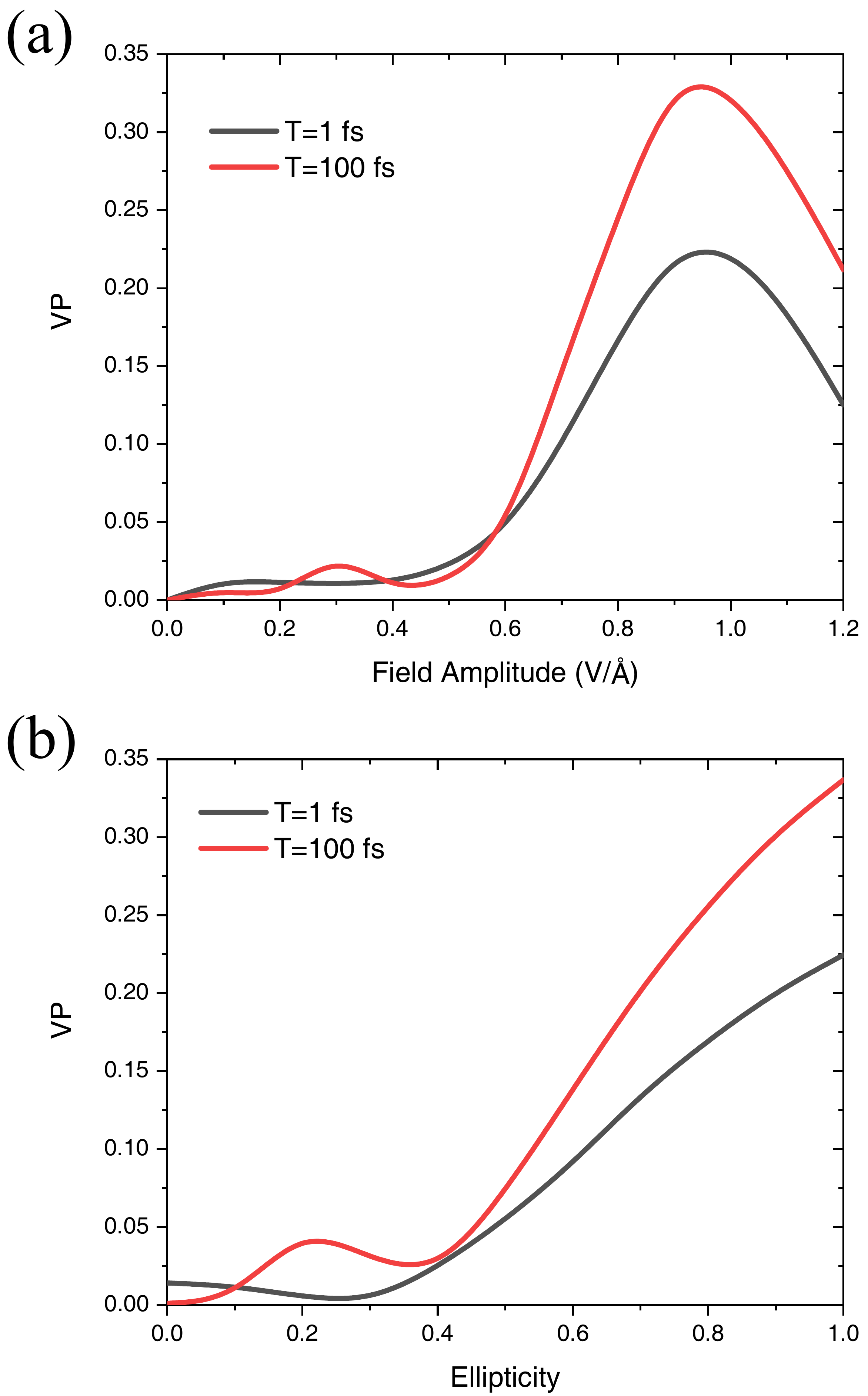}   \caption{ (a) VP as a function of the field amplitude for a circular pulse. Two relaxation times T=1 fs (red line) and T=100 fs (black line). VP increases monotonically with field and reaches to a maximum of ~35 percent (red), and ~22 percent (black, fast dephasing and decoherence) at ${F_0} = 0.9\,{\rm{V/\AA}}$, before sharply falling at higher fields. (b) VP versus pulse ellipticity with field amplitude ${F_0} = 0.9\,{\rm{V/\AA}}$. Similarly, red and black correspond to T=1 and 100 fs, respectively. } 
	\label{Fig_Combined_Comparison_T_1_100} 
\end{figure}

We further look into the influence of the dephasing time, T, on the field amplitude ($F_0$) and ellipticity ($\varepsilon$) dependence of the VP. Fig. \ref{Fig_Combined_Comparison_T_1_100} plots VP as a function of amplitude of the pulse $F_0$ (a), and versus curvature of the pulse $\varepsilon$ (b). Two cases of relaxation time have been examined: The black line correspond to a superfast decaying time (T=1 fs), and the red line correspond to the relatively slow relaxation (T=100 fs). the pulse length is ~5 fs. the fast decay rate suppress the maximum VP from 35 percent to approximately 20 percent, while the general behavior of the VP remains roughly the same as the case with no or slow relaxation.

\begin{figure*} 
	\includegraphics[width=1.0\textwidth]{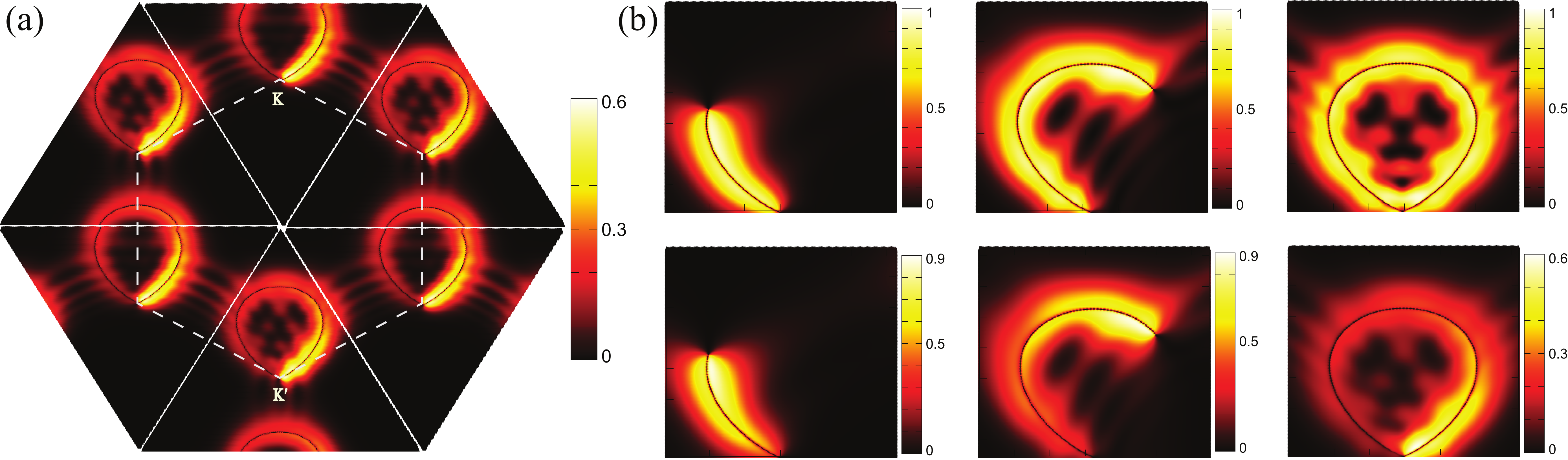}
	\caption{[Color online] (a) The influence of fast relaxation (T=2 fs) on the CB population distribution at the end of the pulse for circular waveform (to be compared with Fig. \ref{Fig_ExtendedBZ_T_100_Hex} with T=100 fs). Compared to Fig. \ref{Fig_ExtendedBZ_T_100_Hex} the maximum population reduces from 1 to 0.6. Also, the CB population deforms along the electron trajectory as time progresses arising to an  asymmetric population distribution (part b).    } 
	\label{Fig_combined_ExtendedBZ_T_2fs_TimeEvolution}
\end{figure*}

In Fig \ref{Fig_combined_ExtendedBZ_T_2fs_TimeEvolution}(a) the impact of relaxation processes on the excitation distribution of graphene is illustrated at the end of the single-cycle circular pulse for decoherence time T=2 fs. The triangular borderline seperating the proportion of $\rm{K}$ and $\rm{K'}$ valleys for calculating ${n_{\rm{K}}}$ and ${n_{\rm{K'}}}$  in Eq. \ref{Eq:NkNKprime} are indicated in the density plot of CB population distribution. The VP, which is calculated by Eq. \ref{Eq:VP} is the total population over the triangular surface of K subtracted from the $\rm{K'}$-triangle, normalized to 1.

 \par

\begin{figure}   
	\centering   
	\includegraphics[width=.95\columnwidth,keepaspectratio ]{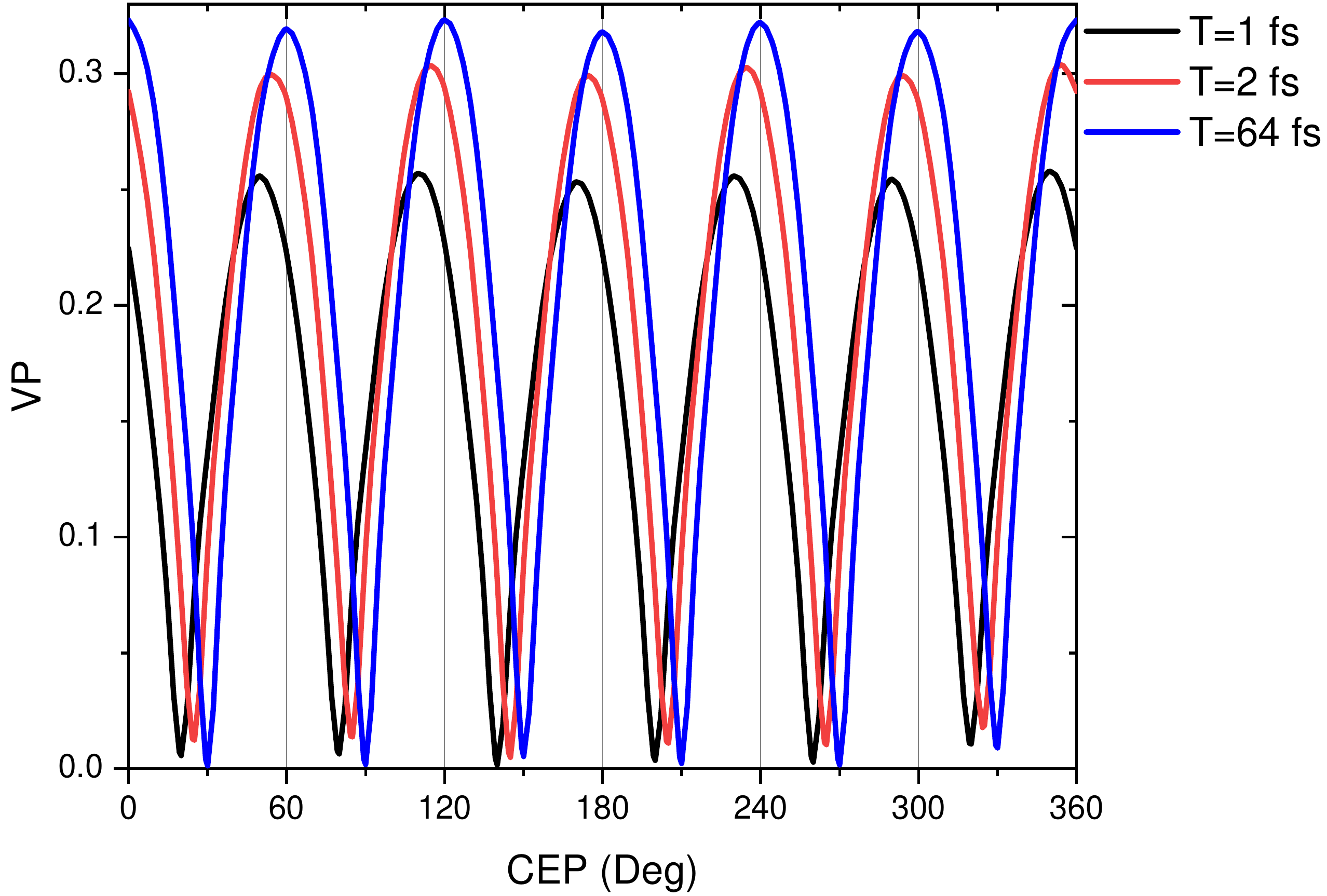}   \caption{ Exhibits VP versus the $\phi _{\rm{CEP}}$ for three different relaxation times. For all case scenarios, the VP periodically modulates with on and off periodicity of 60 degree.  } 
	\label{Fig_VP_CEP} 
\end{figure}

The fast electron scattering impacts on the magnitude of electron wave packet as well as its phase information, giving rise to an asymmetric CB distribution with maximum amplitude of 0.6.  In Fig. \ref{Fig_combined_ExtendedBZ_T_2fs_TimeEvolution}(b) we compare the time evolution of the population formation close to the Dirac point for T=2 fs (lower panel) with the case of long relaxation time T=100 fs (upper panel). The corresponding extended BZ distribution of T=100 is depicted in Fig. \ref{Fig_ExtendedBZ_T_100_Hex}. The maximum population for T=2 fs reduces to 0.6, in contrast to the case for T=100 fs where population maxes out to 1. Besides, the fast relaxation time partially smears the excitation distribution along the cyclic path of electron.

\subsection{\label{Sec:CEP} Carrier Envelope Phase}

We have seen in section \ref{Sec:Field_amplitude} that the VP can be optimally controlled by the amplitude of the laser field. In fact, a threshold of $~0.5  {\rm{V/\AA}}$ is observed in Fig. \ref{Fig_Combined_VP_vs_F_ECC}(a), below which there is no VP irrespective of the polarization state and laser waveform. Above this field threshold, the single cycle pulse exhibits dissimilar behavior for different ellipticities. Likewise, VP behaves in a different way as a function  of ellipticity above and below the $\varepsilon \simeq 0.5 $ threshold (see Fig. \ref{Fig_Combined_VP_vs_F_ECC}b).
 
In Fig. \ref{Fig_VP_CEP} the carrier envelope phase (${\phi _{{\rm{CEP}}}}$ in Eq. \ref{Eq:VectorPotential}) dependence of VP is plotted for three different relaxation times: T = 1, 2, and 100 fs. The case of a circularly polarized pulse is considered with amplitude ${F_0} = 0.9\,{\rm{V/\AA}}$ of the electric field corresponding to the optimum VP observed in Sec. \ref{Sec:Field_amplitude}, where ${\phi _{{\rm{CEP}}}}$ was set to zero. We extend over the full range of CEP from 0 to $2\pi $. 

Fig. \ref{Fig_VP_CEP} reveals that the VP is highly sensitive to the pulse orientation with respect to the graphene sheet which is controlled by the CEP, ${\phi _{{\rm{CEP}}}}$ in Eq. \ref{Eq:VectorPotential}. The VP switches on and off depending on the CEP angle with periodicity of ${60^ \circ }$. Such a periodic behavior is due to the fact that graphene belongs to the Symmorphic ${\rm{C}}_{6v}$ point group with ${60^ \circ }$ rotational symmetry in real and reciprocal space.

Another important physical point which is drawn from Fig. \ref{Fig_VP_CEP} is the shift of VP with respect to the relaxation time. Such a shift can be understood by operating a unitary transformation on elements of the density matrix in Eq. \ref{Eq:Rho_Dot} as ${\rho ^{ij}} \mapsto {{\tilde \rho }^{ij}}{e^{-\gamma t}}$. Subsequently, we find

\begin{equation}
\label{DensityMatrix-shift}
{\begin{array}{*{20}{l}}
	{\dot \tilde \rho _{{\bf{k}}(t)}^{cv} = \frac{ie}{\hbar }{\bf{F}}(t) \cdot {\bf{Q}}_{{\bf{k}}(t)}^{cv}\left[ {\tilde \rho _{{\bf{k}}(t)}^{cc} - \tilde \rho _{{\bf{k}}(t)}^{vv}} \right]}\\
	{\dot \tilde \rho _{{\bf{k}}(t)}^{cc} = 2{\rm{Re}}\left[ {\frac{ie}{\hbar }{\bf{F}}(t) \cdot {\bf{Q}}_{{\bf{k}}(t)}^{cv*}\tilde \rho _{{\bf{k}}(t)}^{cv}} \right]}
	\end{array}}
\end{equation}

Thus in Eq. \ref{Eq:NkNKprime} we can transform the relaxation rate to the exponent. Since the carrier envelope phase acting as a rotation operator on the  electron trajectory in reciprocal space, under the influence of the relaxation processes, acts like a shift vector on the density matrix elements. 



\section{Conclusions}
The principal challenge in the development of valleytronics is to lift the valley degeneracy of charge carriers in a controlled way. The ability to exploit valley polarization (VP) has been rather limited until the recent emergence of the 2D materials. 2D materials with honeycomb structures preeminent by graphene offer a combination of properties not obtainable from conventional thin-film materials.

Since the advent of graphene, various possibilities have been envisioned and explored in it for photonic, plasmonic and optoelectronic devices. However, the generation and detection of valley polarization and feasibility of valleytronics in pristine graphene has been elusive due to the presence of inversion symmetry. In this article, we have proposed a potential prototype of valleytronic device based on monolayer graphene. we have characterized the optimality condition of valley polarization and investigated the parameters upon which we can control the carriers in graphene and store data via valley degree of freedom. 

We have shown the the valley dependent Berry phase results in a valley-contrasting population and carrier transport in pristine graphene. Exciting graphene with sub- 10 femtosecond light pulse creates nonequilibrium charge states in a highly nonlinear fashion. Nonlinear properties are the basis of functional optical devices, as they enable functions such as ultrafast modulation and control, and optical gain.

In fact, driving a system by strong coherent fields break energy-momentum conservation law, and nonadiabatic geometric effects dominates the dynamical processes. Hence, the perturbative optical phenomena which is induced by the absorption and emission of photons is replaced by the nonadiabatic geometric effects. Respectively, a  nonperturbative replica for the common selection rule is set off with the elemental of symmetry, chirality and topology in the optically allowed and forbidden transitions.

According to our results, a circularly polarized pulse with a single-cycle carrier generates a Valley population with efficiency as high as 35$\%$ which is substantial and disprove the general conception that the valley degeneracy can hardly be lifted in gapless materials.  
Moreover, the VP is susceptible to the orientation of the laser waveform with respect to the graphene sheet. VP modulates periodically from its maximum value to almost zero depending on the CEP angle with a periodicity of ${60^ \circ }$, indicating a clear-cut and robust modulation of the VP. This periodic switching of VP with respect to CEP is understandable considering the fact that graphene belongs to the Symmorphic ${\rm{C}}_{6v}$ point group with ${60^ \circ }$ rotational symmetry in real and reciprocal space. 
We further investigate the effect of relaxation on the resultant VP. Notably, even for an ultrafast decoherence and dephasing time of 1 fs, we expect to have more than twenty percent of the valley-contrasting excitation.

The predicted induction of the lightwave valley polarization with a few-cycles of optical lasers in graphene could be useful for valleytronics applications, and development of electronic devices such as valley-polarized optoelectronic emitters, valley optical interconnects, and ultrafast data storage with utmost reliability and robustness.\\

\begin{acknowledgments}

We would like to thank Alexandra Landsman, Mark Stockman, Vadym Apalkov and Lisa Ortmann for fruitful discussions.\\

\end{acknowledgments}

\def\bibsection{\section*{REFERENCES}}
\bibliographystyle{apsrev4-1}
\bibliography{references_Graphene_ValleyPolarization}

\begin{thebibliography}{48}%
\makeatletter
\providecommand \@ifxundefined [1]{%
 \@ifx{#1\undefined}
}%
\providecommand \@ifnum [1]{%
 \ifnum #1\expandafter \@firstoftwo
 \else \expandafter \@secondoftwo
 \fi
}%
\providecommand \@ifx [1]{%
 \ifx #1\expandafter \@firstoftwo
 \else \expandafter \@secondoftwo
 \fi
}%
\providecommand \natexlab [1]{#1}%
\providecommand \enquote  [1]{``#1''}%
\providecommand \bibnamefont  [1]{#1}%
\providecommand \bibfnamefont [1]{#1}%
\providecommand \citenamefont [1]{#1}%
\providecommand \href@noop [0]{\@secondoftwo}%
\providecommand \href [0]{\begingroup \@sanitize@url \@href}%
\providecommand \@href[1]{\@@startlink{#1}\@@href}%
\providecommand \@@href[1]{\endgroup#1\@@endlink}%
\providecommand \@sanitize@url [0]{\catcode `\\12\catcode `\$12\catcode
  `\&12\catcode `\#12\catcode `\^12\catcode `\_12\catcode `\%12\relax}%
\providecommand \@@startlink[1]{}%
\providecommand \@@endlink[0]{}%
\providecommand \url  [0]{\begingroup\@sanitize@url \@url }%
\providecommand \@url [1]{\endgroup\@href {#1}{\urlprefix }}%
\providecommand \urlprefix  [0]{URL }%
\providecommand \Eprint [0]{\href }%
\providecommand \doibase [0]{http://dx.doi.org/}%
\providecommand \selectlanguage [0]{\@gobble}%
\providecommand \bibinfo  [0]{\@secondoftwo}%
\providecommand \bibfield  [0]{\@secondoftwo}%
\providecommand \translation [1]{[#1]}%
\providecommand \BibitemOpen [0]{}%
\providecommand \bibitemStop [0]{}%
\providecommand \bibitemNoStop [0]{.\EOS\space}%
\providecommand \EOS [0]{\spacefactor3000\relax}%
\providecommand \BibitemShut  [1]{\csname bibitem#1\endcsname}%
\let\auto@bib@innerbib\@empty
\bibitem [{\citenamefont {Xiao}\ \emph {et~al.}(2010)\citenamefont {Xiao},
  \citenamefont {Chang},\ and\ \citenamefont {Niu}}]{Xiao2010}%
  \BibitemOpen
  \bibfield  {author} {\bibinfo {author} {\bibfnamefont {D.}~\bibnamefont
  {Xiao}}, \bibinfo {author} {\bibfnamefont {M.~C.}\ \bibnamefont {Chang}}, \
  and\ \bibinfo {author} {\bibfnamefont {Q.}~\bibnamefont {Niu}},\ }\href
  {\doibase 10.1103/RevModPhys.82.1959} {\bibfield  {journal} {\bibinfo
  {journal} {Reviews of Modern Physics}\ }\textbf {\bibinfo {volume} {82}},\
  \bibinfo {pages} {1959} (\bibinfo {year} {2010})},\ \Eprint
  {http://arxiv.org/abs/0907.2021} {0907.2021} \BibitemShut {NoStop}%
\bibitem [{\citenamefont {Qi}\ and\ \citenamefont {Zhang}(2011)}]{Qi2011}%
  \BibitemOpen
  \bibfield  {author} {\bibinfo {author} {\bibfnamefont {X.~L.}\ \bibnamefont
  {Qi}}\ and\ \bibinfo {author} {\bibfnamefont {S.~C.}\ \bibnamefont {Zhang}},\
  }\href {\doibase 10.1103/RevModPhys.83.1057} {\bibfield  {journal} {\bibinfo
  {journal} {Reviews of Modern Physics}\ }\textbf {\bibinfo {volume} {83}},\
  \bibinfo {pages} {1057} (\bibinfo {year} {2011})},\ \Eprint
  {http://arxiv.org/abs/1008.2026} {1008.2026} \BibitemShut {NoStop}%
\bibitem [{\citenamefont {Schaibley}\ \emph {et~al.}(2016)\citenamefont
  {Schaibley}, \citenamefont {Yu}, \citenamefont {Clark}, \citenamefont
  {Rivera}, \citenamefont {Ross}, \citenamefont {Seyler}, \citenamefont {Yao},\
  and\ \citenamefont {Xu}}]{Schaibley2016}%
  \BibitemOpen
  \bibfield  {author} {\bibinfo {author} {\bibfnamefont {J.~R.}\ \bibnamefont
  {Schaibley}}, \bibinfo {author} {\bibfnamefont {H.}~\bibnamefont {Yu}},
  \bibinfo {author} {\bibfnamefont {G.}~\bibnamefont {Clark}}, \bibinfo
  {author} {\bibfnamefont {P.}~\bibnamefont {Rivera}}, \bibinfo {author}
  {\bibfnamefont {J.~S.}\ \bibnamefont {Ross}}, \bibinfo {author}
  {\bibfnamefont {K.~L.}\ \bibnamefont {Seyler}}, \bibinfo {author}
  {\bibfnamefont {W.}~\bibnamefont {Yao}}, \ and\ \bibinfo {author}
  {\bibfnamefont {X.}~\bibnamefont {Xu}},\ }\href {\doibase
  10.1038/natrevmats.2016.55} {\bibfield  {journal} {\bibinfo  {journal}
  {Nature Reviews Materials}\ }\textbf {\bibinfo {volume} {1}},\ \bibinfo
  {pages} {16055} (\bibinfo {year} {2016})}\BibitemShut {NoStop}%
\bibitem [{\citenamefont {Kelardeh}\ \emph {et~al.}(2016)\citenamefont
  {Kelardeh}, \citenamefont {Apalkov},\ and\ \citenamefont
  {Stockman}}]{Kelardeh2016_Attosecond}%
  \BibitemOpen
  \bibfield  {author} {\bibinfo {author} {\bibfnamefont {H.~K.}\ \bibnamefont
  {Kelardeh}}, \bibinfo {author} {\bibfnamefont {V.}~\bibnamefont {Apalkov}}, \
  and\ \bibinfo {author} {\bibfnamefont {M.~I.}\ \bibnamefont {Stockman}},\
  }\href {\doibase 10.1103/PhysRevB.93.155434} {\bibfield  {journal} {\bibinfo
  {journal} {Physical Review B}\ }\textbf {\bibinfo {volume} {93}},\ \bibinfo
  {pages} {155434} (\bibinfo {year} {2016})}\BibitemShut {NoStop}%
\bibitem [{\citenamefont {Khanikaev}\ and\ \citenamefont
  {Shvets}(2017)}]{Khanikaev2017}%
  \BibitemOpen
  \bibfield  {author} {\bibinfo {author} {\bibfnamefont {A.~B.}\ \bibnamefont
  {Khanikaev}}\ and\ \bibinfo {author} {\bibfnamefont {G.}~\bibnamefont
  {Shvets}},\ }\href {\doibase 10.1038/s41566-017-0048-5} {\bibfield  {journal}
  {\bibinfo  {journal} {Nature Photonics}\ }\textbf {\bibinfo {volume} {11}},\
  \bibinfo {pages} {763} (\bibinfo {year} {2017})}\BibitemShut {NoStop}%
\bibitem [{\citenamefont {Vitale}\ \emph {et~al.}(2018)\citenamefont {Vitale},
  \citenamefont {Nezich}, \citenamefont {Varghese}, \citenamefont {Kim},
  \citenamefont {Gedik}, \citenamefont {Jarillo-Herrero}, \citenamefont
  {Xiao},\ and\ \citenamefont {Rothschild}}]{Vitale2018}%
  \BibitemOpen
  \bibfield  {author} {\bibinfo {author} {\bibfnamefont {S.~A.}\ \bibnamefont
  {Vitale}}, \bibinfo {author} {\bibfnamefont {D.}~\bibnamefont {Nezich}},
  \bibinfo {author} {\bibfnamefont {J.~O.}\ \bibnamefont {Varghese}}, \bibinfo
  {author} {\bibfnamefont {P.}~\bibnamefont {Kim}}, \bibinfo {author}
  {\bibfnamefont {N.}~\bibnamefont {Gedik}}, \bibinfo {author} {\bibfnamefont
  {P.}~\bibnamefont {Jarillo-Herrero}}, \bibinfo {author} {\bibfnamefont
  {D.}~\bibnamefont {Xiao}}, \ and\ \bibinfo {author} {\bibfnamefont
  {M.}~\bibnamefont {Rothschild}},\ }\href {\doibase 10.1002/smll.201870172}
  {\bibfield  {journal} {\bibinfo  {journal} {Small}\ }\textbf {\bibinfo
  {volume} {14}},\ \bibinfo {pages} {1870172} (\bibinfo {year}
  {2018})}\BibitemShut {NoStop}%
\bibitem [{\citenamefont {Jiménez-Galán}\ \emph {et~al.}(2020)\citenamefont
  {Jiménez-Galán}, \citenamefont {Silva}, \citenamefont {Smirnova},\ and\
  \citenamefont {Ivanov}}]{Galan2019}%
  \BibitemOpen
  \bibfield  {author} {\bibinfo {author} {\bibfnamefont {A.}~\bibnamefont
  {Jiménez-Galán}}, \bibinfo {author} {\bibfnamefont {R.~E.~F.}\ \bibnamefont
  {Silva}}, \bibinfo {author} {\bibfnamefont {O.}~\bibnamefont {Smirnova}}, \
  and\ \bibinfo {author} {\bibfnamefont {M.}~\bibnamefont {Ivanov}},\ }\href
  {\doibase 10.1038/s41566-020-00717-3} {\bibfield  {journal} {\bibinfo
  {journal} {Nature Photonics}\ }\textbf {\bibinfo {volume} {14}},\ \bibinfo
  {pages} {728} (\bibinfo {year} {2020})}\BibitemShut {NoStop}%
\bibitem [{\citenamefont {Ozawa}\ \emph {et~al.}(2019)\citenamefont {Ozawa},
  \citenamefont {Price}, \citenamefont {Amo}, \citenamefont {Goldman},
  \citenamefont {Hafezi}, \citenamefont {Lu}, \citenamefont {Rechtsman},
  \citenamefont {Schuster}, \citenamefont {Simon}, \citenamefont {Zilberberg},\
  and\ \citenamefont {Carusotto}}]{Hafezi2019}%
  \BibitemOpen
  \bibfield  {author} {\bibinfo {author} {\bibfnamefont {T.}~\bibnamefont
  {Ozawa}}, \bibinfo {author} {\bibfnamefont {H.~M.}\ \bibnamefont {Price}},
  \bibinfo {author} {\bibfnamefont {A.}~\bibnamefont {Amo}}, \bibinfo {author}
  {\bibfnamefont {N.}~\bibnamefont {Goldman}}, \bibinfo {author} {\bibfnamefont
  {M.}~\bibnamefont {Hafezi}}, \bibinfo {author} {\bibfnamefont
  {L.}~\bibnamefont {Lu}}, \bibinfo {author} {\bibfnamefont {M.~C.}\
  \bibnamefont {Rechtsman}}, \bibinfo {author} {\bibfnamefont {D.}~\bibnamefont
  {Schuster}}, \bibinfo {author} {\bibfnamefont {J.}~\bibnamefont {Simon}},
  \bibinfo {author} {\bibfnamefont {O.}~\bibnamefont {Zilberberg}}, \ and\
  \bibinfo {author} {\bibfnamefont {I.}~\bibnamefont {Carusotto}},\ }\href
  {\doibase ARTN 015006 10.1103/RevModPhys.91.015006} {\bibfield  {journal}
  {\bibinfo  {journal} {Reviews of Modern Physics}\ }\textbf {\bibinfo {volume}
  {91}},\ \bibinfo {pages} {015006} (\bibinfo {year} {2019})}\BibitemShut
  {NoStop}%
\bibitem [{\citenamefont {Song}\ \emph {et~al.}(2017)\citenamefont {Song},
  \citenamefont {Li}, \citenamefont {Wang}, \citenamefont {Bai}, \citenamefont
  {Wang}, \citenamefont {Du}, \citenamefont {Liu}, \citenamefont {Wang},
  \citenamefont {Han}, \citenamefont {Yang}, \citenamefont {Liu}, \citenamefont
  {Lu}, \citenamefont {Fang},\ and\ \citenamefont {Yang}}]{Song2017}%
  \BibitemOpen
  \bibfield  {author} {\bibinfo {author} {\bibfnamefont {Z.}~\bibnamefont
  {Song}}, \bibinfo {author} {\bibfnamefont {Z.}~\bibnamefont {Li}}, \bibinfo
  {author} {\bibfnamefont {H.}~\bibnamefont {Wang}}, \bibinfo {author}
  {\bibfnamefont {X.}~\bibnamefont {Bai}}, \bibinfo {author} {\bibfnamefont
  {W.}~\bibnamefont {Wang}}, \bibinfo {author} {\bibfnamefont {H.}~\bibnamefont
  {Du}}, \bibinfo {author} {\bibfnamefont {S.}~\bibnamefont {Liu}}, \bibinfo
  {author} {\bibfnamefont {C.}~\bibnamefont {Wang}}, \bibinfo {author}
  {\bibfnamefont {J.}~\bibnamefont {Han}}, \bibinfo {author} {\bibfnamefont
  {Y.}~\bibnamefont {Yang}}, \bibinfo {author} {\bibfnamefont {Z.}~\bibnamefont
  {Liu}}, \bibinfo {author} {\bibfnamefont {J.}~\bibnamefont {Lu}}, \bibinfo
  {author} {\bibfnamefont {Z.}~\bibnamefont {Fang}}, \ and\ \bibinfo {author}
  {\bibfnamefont {J.}~\bibnamefont {Yang}},\ }\href {\doibase
  10.1021/acs.nanolett.7b00271} {\bibfield  {journal} {\bibinfo  {journal}
  {Nano Letters}\ }\textbf {\bibinfo {volume} {17}},\ \bibinfo {pages} {2079}
  (\bibinfo {year} {2017})}\BibitemShut {NoStop}%
\bibitem [{\citenamefont {Isberg}\ \emph {et~al.}(2013)\citenamefont {Isberg},
  \citenamefont {Gabrysch}, \citenamefont {Hammersberg}, \citenamefont {Majdi},
  \citenamefont {Kovi},\ and\ \citenamefont {Twitchen}}]{Isberg2013}%
  \BibitemOpen
  \bibfield  {author} {\bibinfo {author} {\bibfnamefont {J.}~\bibnamefont
  {Isberg}}, \bibinfo {author} {\bibfnamefont {M.}~\bibnamefont {Gabrysch}},
  \bibinfo {author} {\bibfnamefont {J.}~\bibnamefont {Hammersberg}}, \bibinfo
  {author} {\bibfnamefont {S.}~\bibnamefont {Majdi}}, \bibinfo {author}
  {\bibfnamefont {K.~K.}\ \bibnamefont {Kovi}}, \ and\ \bibinfo {author}
  {\bibfnamefont {D.~J.}\ \bibnamefont {Twitchen}},\ }\href {\doibase
  10.1038/nmat3694} {\bibfield  {journal} {\bibinfo  {journal} {Nature
  Materials}\ }\textbf {\bibinfo {volume} {12}},\ \bibinfo {pages} {760}
  (\bibinfo {year} {2013})}\BibitemShut {NoStop}%
\bibitem [{\citenamefont {Cao}\ \emph {et~al.}(2012)\citenamefont {Cao},
  \citenamefont {Wang}, \citenamefont {Han}, \citenamefont {Ye}, \citenamefont
  {Zhu}, \citenamefont {Shi}, \citenamefont {Niu}, \citenamefont {Tan},
  \citenamefont {Wang}, \citenamefont {Liu},\ and\ \citenamefont
  {Feng}}]{Cao2012}%
  \BibitemOpen
  \bibfield  {author} {\bibinfo {author} {\bibfnamefont {T.}~\bibnamefont
  {Cao}}, \bibinfo {author} {\bibfnamefont {G.}~\bibnamefont {Wang}}, \bibinfo
  {author} {\bibfnamefont {W.}~\bibnamefont {Han}}, \bibinfo {author}
  {\bibfnamefont {H.}~\bibnamefont {Ye}}, \bibinfo {author} {\bibfnamefont
  {C.}~\bibnamefont {Zhu}}, \bibinfo {author} {\bibfnamefont {J.}~\bibnamefont
  {Shi}}, \bibinfo {author} {\bibfnamefont {Q.}~\bibnamefont {Niu}}, \bibinfo
  {author} {\bibfnamefont {P.}~\bibnamefont {Tan}}, \bibinfo {author}
  {\bibfnamefont {E.}~\bibnamefont {Wang}}, \bibinfo {author} {\bibfnamefont
  {B.}~\bibnamefont {Liu}}, \ and\ \bibinfo {author} {\bibfnamefont
  {J.}~\bibnamefont {Feng}},\ }\href {\doibase 10.1038/ncomms1882} {\bibfield
  {journal} {\bibinfo  {journal} {Nature Communications}\ }\textbf {\bibinfo
  {volume} {3}},\ \bibinfo {pages} {887} (\bibinfo {year} {2012})}\BibitemShut
  {NoStop}%
\bibitem [{\citenamefont {Mak}\ \emph {et~al.}(2012)\citenamefont {Mak},
  \citenamefont {He}, \citenamefont {Shan},\ and\ \citenamefont
  {Heinz}}]{Mak2012}%
  \BibitemOpen
  \bibfield  {author} {\bibinfo {author} {\bibfnamefont {K.~F.}\ \bibnamefont
  {Mak}}, \bibinfo {author} {\bibfnamefont {K.}~\bibnamefont {He}}, \bibinfo
  {author} {\bibfnamefont {J.}~\bibnamefont {Shan}}, \ and\ \bibinfo {author}
  {\bibfnamefont {T.~F.}\ \bibnamefont {Heinz}},\ }\href {\doibase
  10.1038/nnano.2012.96} {\bibfield  {journal} {\bibinfo  {journal} {Nature
  Nanotechnology}\ }\textbf {\bibinfo {volume} {7}},\ \bibinfo {pages} {494}
  (\bibinfo {year} {2012})},\ \Eprint {http://arxiv.org/abs/1205.1822}
  {1205.1822} \BibitemShut {NoStop}%
\bibitem [{\citenamefont {Jones}\ \emph {et~al.}(2013)\citenamefont {Jones},
  \citenamefont {Yu}, \citenamefont {Ghimire}, \citenamefont {Wu},
  \citenamefont {Aivazian}, \citenamefont {Ross}, \citenamefont {Zhao},
  \citenamefont {Yan}, \citenamefont {Mandrus}, \citenamefont {Xiao},
  \citenamefont {Yao},\ and\ \citenamefont {Xu}}]{Jones2013}%
  \BibitemOpen
  \bibfield  {author} {\bibinfo {author} {\bibfnamefont {A.~M.}\ \bibnamefont
  {Jones}}, \bibinfo {author} {\bibfnamefont {H.}~\bibnamefont {Yu}}, \bibinfo
  {author} {\bibfnamefont {N.~J.}\ \bibnamefont {Ghimire}}, \bibinfo {author}
  {\bibfnamefont {S.}~\bibnamefont {Wu}}, \bibinfo {author} {\bibfnamefont
  {G.}~\bibnamefont {Aivazian}}, \bibinfo {author} {\bibfnamefont {J.~S.}\
  \bibnamefont {Ross}}, \bibinfo {author} {\bibfnamefont {B.}~\bibnamefont
  {Zhao}}, \bibinfo {author} {\bibfnamefont {J.}~\bibnamefont {Yan}}, \bibinfo
  {author} {\bibfnamefont {D.~G.}\ \bibnamefont {Mandrus}}, \bibinfo {author}
  {\bibfnamefont {D.}~\bibnamefont {Xiao}}, \bibinfo {author} {\bibfnamefont
  {W.}~\bibnamefont {Yao}}, \ and\ \bibinfo {author} {\bibfnamefont
  {X.}~\bibnamefont {Xu}},\ }\href {\doibase 10.1038/nnano.2013.151} {\bibfield
   {journal} {\bibinfo  {journal} {Nature Nanotechnology}\ }\textbf {\bibinfo
  {volume} {8}},\ \bibinfo {pages} {634} (\bibinfo {year} {2013})}\BibitemShut
  {NoStop}%
\bibitem [{\citenamefont {Yoshikawa}\ \emph {et~al.}(2019)\citenamefont
  {Yoshikawa}, \citenamefont {Nagai}, \citenamefont {Uchida}, \citenamefont
  {Takaguchi}, \citenamefont {Sasaki}, \citenamefont {Miyata},\ and\
  \citenamefont {Tanaka}}]{Yoshikawa2019}%
  \BibitemOpen
  \bibfield  {author} {\bibinfo {author} {\bibfnamefont {N.}~\bibnamefont
  {Yoshikawa}}, \bibinfo {author} {\bibfnamefont {K.}~\bibnamefont {Nagai}},
  \bibinfo {author} {\bibfnamefont {K.}~\bibnamefont {Uchida}}, \bibinfo
  {author} {\bibfnamefont {Y.}~\bibnamefont {Takaguchi}}, \bibinfo {author}
  {\bibfnamefont {S.}~\bibnamefont {Sasaki}}, \bibinfo {author} {\bibfnamefont
  {Y.}~\bibnamefont {Miyata}}, \ and\ \bibinfo {author} {\bibfnamefont
  {K.}~\bibnamefont {Tanaka}},\ }\href {\doibase 10.1038/s41467-019-11697-6}
  {\bibfield  {journal} {\bibinfo  {journal} {Nature Communications}\ }\textbf
  {\bibinfo {volume} {10}},\ \bibinfo {pages} {3709} (\bibinfo {year}
  {2019})}\BibitemShut {NoStop}%
\bibitem [{\citenamefont {Zeng}\ \emph {et~al.}(2012)\citenamefont {Zeng},
  \citenamefont {Dai}, \citenamefont {Yao}, \citenamefont {Xiao},\ and\
  \citenamefont {Cui}}]{Zeng2012}%
  \BibitemOpen
  \bibfield  {author} {\bibinfo {author} {\bibfnamefont {H.}~\bibnamefont
  {Zeng}}, \bibinfo {author} {\bibfnamefont {J.}~\bibnamefont {Dai}}, \bibinfo
  {author} {\bibfnamefont {W.}~\bibnamefont {Yao}}, \bibinfo {author}
  {\bibfnamefont {D.}~\bibnamefont {Xiao}}, \ and\ \bibinfo {author}
  {\bibfnamefont {X.}~\bibnamefont {Cui}},\ }\href {\doibase
  10.1038/nnano.2012.95} {\bibfield  {journal} {\bibinfo  {journal} {Nature
  Nanotechnology}\ }\textbf {\bibinfo {volume} {7}},\ \bibinfo {pages} {490}
  (\bibinfo {year} {2012})}\BibitemShut {NoStop}%
\bibitem [{\citenamefont {Ye}\ \emph {et~al.}(2017)\citenamefont {Ye},
  \citenamefont {Sun},\ and\ \citenamefont {Heinz}}]{Heinz2017}%
  \BibitemOpen
  \bibfield  {author} {\bibinfo {author} {\bibfnamefont {Z.}~\bibnamefont
  {Ye}}, \bibinfo {author} {\bibfnamefont {D.}~\bibnamefont {Sun}}, \ and\
  \bibinfo {author} {\bibfnamefont {T.~F.}\ \bibnamefont {Heinz}},\ }\href
  {\doibase 10.1038/nphys3891} {\bibfield  {journal} {\bibinfo  {journal}
  {Nature Physics}\ }\textbf {\bibinfo {volume} {13}},\ \bibinfo {pages} {26}
  (\bibinfo {year} {2017})}\BibitemShut {NoStop}%
\bibitem [{\citenamefont {Langer}\ \emph {et~al.}(2018)\citenamefont {Langer},
  \citenamefont {Schmid}, \citenamefont {Schlauderer}, \citenamefont {Gmitra},
  \citenamefont {Fabian}, \citenamefont {Nagler}, \citenamefont
  {Sch{\"{u}}ller}, \citenamefont {Korn}, \citenamefont {Hawkins},
  \citenamefont {Steiner}, \citenamefont {Huttner}, \citenamefont {Koch},
  \citenamefont {Kira},\ and\ \citenamefont {Huber}}]{Langer2018}%
  \BibitemOpen
  \bibfield  {author} {\bibinfo {author} {\bibfnamefont {F.}~\bibnamefont
  {Langer}}, \bibinfo {author} {\bibfnamefont {C.~P.}\ \bibnamefont {Schmid}},
  \bibinfo {author} {\bibfnamefont {S.}~\bibnamefont {Schlauderer}}, \bibinfo
  {author} {\bibfnamefont {M.}~\bibnamefont {Gmitra}}, \bibinfo {author}
  {\bibfnamefont {J.}~\bibnamefont {Fabian}}, \bibinfo {author} {\bibfnamefont
  {P.}~\bibnamefont {Nagler}}, \bibinfo {author} {\bibfnamefont
  {C.}~\bibnamefont {Sch{\"{u}}ller}}, \bibinfo {author} {\bibfnamefont
  {T.}~\bibnamefont {Korn}}, \bibinfo {author} {\bibfnamefont {P.~G.}\
  \bibnamefont {Hawkins}}, \bibinfo {author} {\bibfnamefont {J.~T.}\
  \bibnamefont {Steiner}}, \bibinfo {author} {\bibfnamefont {U.}~\bibnamefont
  {Huttner}}, \bibinfo {author} {\bibfnamefont {S.~W.}\ \bibnamefont {Koch}},
  \bibinfo {author} {\bibfnamefont {M.}~\bibnamefont {Kira}}, \ and\ \bibinfo
  {author} {\bibfnamefont {R.}~\bibnamefont {Huber}},\ }\href {\doibase
  10.1038/s41586-018-0013-6} {\bibfield  {journal} {\bibinfo  {journal}
  {Nature}\ }\textbf {\bibinfo {volume} {557}},\ \bibinfo {pages} {76}
  (\bibinfo {year} {2018})}\BibitemShut {NoStop}%
\bibitem [{\citenamefont {Salfi}\ \emph {et~al.}(2014)\citenamefont {Salfi},
  \citenamefont {Mol}, \citenamefont {Rahman}, \citenamefont {Klimeck},
  \citenamefont {Simmons}, \citenamefont {Hollenberg},\ and\ \citenamefont
  {Rogge}}]{Salfi2014}%
  \BibitemOpen
  \bibfield  {author} {\bibinfo {author} {\bibfnamefont {J.}~\bibnamefont
  {Salfi}}, \bibinfo {author} {\bibfnamefont {J.~A.}\ \bibnamefont {Mol}},
  \bibinfo {author} {\bibfnamefont {R.}~\bibnamefont {Rahman}}, \bibinfo
  {author} {\bibfnamefont {G.}~\bibnamefont {Klimeck}}, \bibinfo {author}
  {\bibfnamefont {M.~Y.}\ \bibnamefont {Simmons}}, \bibinfo {author}
  {\bibfnamefont {L.~C.~L.}\ \bibnamefont {Hollenberg}}, \ and\ \bibinfo
  {author} {\bibfnamefont {S.}~\bibnamefont {Rogge}},\ }\href {\doibase
  10.1038/nmat3941} {\bibfield  {journal} {\bibinfo  {journal} {Nature
  Materials}\ }\textbf {\bibinfo {volume} {13}},\ \bibinfo {pages} {605}
  (\bibinfo {year} {2014})}\BibitemShut {NoStop}%
\bibitem [{\citenamefont {Zhu}\ \emph {et~al.}(2012)\citenamefont {Zhu},
  \citenamefont {Collaudin}, \citenamefont {Fauqu{\'{e}}}, \citenamefont
  {Kang},\ and\ \citenamefont {Behnia}}]{Zhu2012}%
  \BibitemOpen
  \bibfield  {author} {\bibinfo {author} {\bibfnamefont {Z.}~\bibnamefont
  {Zhu}}, \bibinfo {author} {\bibfnamefont {A.}~\bibnamefont {Collaudin}},
  \bibinfo {author} {\bibfnamefont {B.}~\bibnamefont {Fauqu{\'{e}}}}, \bibinfo
  {author} {\bibfnamefont {W.}~\bibnamefont {Kang}}, \ and\ \bibinfo {author}
  {\bibfnamefont {K.}~\bibnamefont {Behnia}},\ }\href {\doibase
  10.1038/nphys2111} {\bibfield  {journal} {\bibinfo  {journal} {Nature
  Physics}\ }\textbf {\bibinfo {volume} {8}},\ \bibinfo {pages} {89} (\bibinfo
  {year} {2012})},\ \Eprint {http://arxiv.org/abs/1109.2774} {1109.2774}
  \BibitemShut {NoStop}%
\bibitem [{\citenamefont {Zhu}\ \emph {et~al.}(2017)\citenamefont {Zhu},
  \citenamefont {Wang}, \citenamefont {Zuo}, \citenamefont {Fauque},
  \citenamefont {McDonald}, \citenamefont {Fuseya},\ and\ \citenamefont
  {Behnia}}]{Zhu_Behnia2017}%
  \BibitemOpen
  \bibfield  {author} {\bibinfo {author} {\bibfnamefont {Z.}~\bibnamefont
  {Zhu}}, \bibinfo {author} {\bibfnamefont {J.}~\bibnamefont {Wang}}, \bibinfo
  {author} {\bibfnamefont {H.}~\bibnamefont {Zuo}}, \bibinfo {author}
  {\bibfnamefont {B.}~\bibnamefont {Fauque}}, \bibinfo {author} {\bibfnamefont
  {R.~D.}\ \bibnamefont {McDonald}}, \bibinfo {author} {\bibfnamefont
  {Y.}~\bibnamefont {Fuseya}}, \ and\ \bibinfo {author} {\bibfnamefont
  {K.}~\bibnamefont {Behnia}},\ }\href {\doibase 10.1038/ncomms15297}
  {\bibfield  {journal} {\bibinfo  {journal} {Nat Commun}\ }\textbf {\bibinfo
  {volume} {8}},\ \bibinfo {pages} {15297} (\bibinfo {year}
  {2017})}\BibitemShut {NoStop}%
\bibitem [{\citenamefont {{Azar Oliaei Motlagh}}\ \emph
  {et~al.}(2019)\citenamefont {{Azar Oliaei Motlagh}}, \citenamefont
  {Nematollahi}, \citenamefont {Apalkov},\ and\ \citenamefont
  {Stockman}}]{Motlagh2019_gapped}%
  \BibitemOpen
  \bibfield  {author} {\bibinfo {author} {\bibfnamefont {S.}~\bibnamefont
  {{Azar Oliaei Motlagh}}}, \bibinfo {author} {\bibfnamefont {F.}~\bibnamefont
  {Nematollahi}}, \bibinfo {author} {\bibfnamefont {V.}~\bibnamefont
  {Apalkov}}, \ and\ \bibinfo {author} {\bibfnamefont {M.~I.}\ \bibnamefont
  {Stockman}},\ }\href {\doibase 10.1103/PhysRevB.100.115431} {\bibfield
  {journal} {\bibinfo  {journal} {Physical Review B}\ }\textbf {\bibinfo
  {volume} {100}},\ \bibinfo {pages} {115431} (\bibinfo {year}
  {2019})}\BibitemShut {NoStop}%
\bibitem [{\citenamefont {Shimazaki}\ \emph {et~al.}(2015)\citenamefont
  {Shimazaki}, \citenamefont {Yamamoto}, \citenamefont {Borzenets},
  \citenamefont {Watanabe}, \citenamefont {Taniguchi},\ and\ \citenamefont
  {Tarucha}}]{Shimazaki2015}%
  \BibitemOpen
  \bibfield  {author} {\bibinfo {author} {\bibfnamefont {Y.}~\bibnamefont
  {Shimazaki}}, \bibinfo {author} {\bibfnamefont {M.}~\bibnamefont {Yamamoto}},
  \bibinfo {author} {\bibfnamefont {I.~V.}\ \bibnamefont {Borzenets}}, \bibinfo
  {author} {\bibfnamefont {K.}~\bibnamefont {Watanabe}}, \bibinfo {author}
  {\bibfnamefont {T.}~\bibnamefont {Taniguchi}}, \ and\ \bibinfo {author}
  {\bibfnamefont {S.}~\bibnamefont {Tarucha}},\ }\href {\doibase
  10.1038/nphys3551
  https://www.nature.com/articles/nphys3551#supplementary-information}
  {\bibfield  {journal} {\bibinfo  {journal} {Nature Physics}\ }\textbf
  {\bibinfo {volume} {11}},\ \bibinfo {pages} {1032} (\bibinfo {year}
  {2015})}\BibitemShut {NoStop}%
\bibitem [{\citenamefont {Sui}\ \emph {et~al.}(2015)\citenamefont {Sui},
  \citenamefont {Chen}, \citenamefont {Ma}, \citenamefont {Shan}, \citenamefont
  {Tian}, \citenamefont {Watanabe}, \citenamefont {Taniguchi}, \citenamefont
  {Jin}, \citenamefont {Yao}, \citenamefont {Xiao},\ and\ \citenamefont
  {Zhang}}]{Sui2015}%
  \BibitemOpen
  \bibfield  {author} {\bibinfo {author} {\bibfnamefont {M.~Q.}\ \bibnamefont
  {Sui}}, \bibinfo {author} {\bibfnamefont {G.~R.}\ \bibnamefont {Chen}},
  \bibinfo {author} {\bibfnamefont {L.~G.}\ \bibnamefont {Ma}}, \bibinfo
  {author} {\bibfnamefont {W.~Y.}\ \bibnamefont {Shan}}, \bibinfo {author}
  {\bibfnamefont {D.}~\bibnamefont {Tian}}, \bibinfo {author} {\bibfnamefont
  {K.}~\bibnamefont {Watanabe}}, \bibinfo {author} {\bibfnamefont
  {T.}~\bibnamefont {Taniguchi}}, \bibinfo {author} {\bibfnamefont {X.~F.}\
  \bibnamefont {Jin}}, \bibinfo {author} {\bibfnamefont {W.}~\bibnamefont
  {Yao}}, \bibinfo {author} {\bibfnamefont {D.}~\bibnamefont {Xiao}}, \ and\
  \bibinfo {author} {\bibfnamefont {Y.~B.}\ \bibnamefont {Zhang}},\ }\href
  {\doibase 10.1038/Nphys3485} {\bibfield  {journal} {\bibinfo  {journal}
  {Nature Physics}\ }\textbf {\bibinfo {volume} {11}},\ \bibinfo {pages} {1027}
  (\bibinfo {year} {2015})}\BibitemShut {NoStop}%
\bibitem [{\citenamefont {Kumar}\ \emph {et~al.}(2020)\citenamefont {Kumar},
  \citenamefont {Herath}, \citenamefont {Apalkov},\ and\ \citenamefont
  {Stockman}}]{Kumar2020}%
  \BibitemOpen
  \bibfield  {author} {\bibinfo {author} {\bibfnamefont {P.}~\bibnamefont
  {Kumar}}, \bibinfo {author} {\bibfnamefont {T.~M.}\ \bibnamefont {Herath}},
  \bibinfo {author} {\bibfnamefont {V.}~\bibnamefont {Apalkov}}, \ and\
  \bibinfo {author} {\bibfnamefont {M.~I.}\ \bibnamefont {Stockman}},\
  }\href@noop {} {\bibfield  {journal} {\bibinfo  {journal} {arXiv preprint
  arXiv:2007.13480}\ } (\bibinfo {year} {2020})}\BibitemShut {NoStop}%
\bibitem [{\citenamefont {Yankowitz}\ \emph {et~al.}(2012)\citenamefont
  {Yankowitz}, \citenamefont {Xue}, \citenamefont {Cormode}, \citenamefont
  {Sanchez-Yamagishi}, \citenamefont {Watanabe}, \citenamefont {Taniguchi},
  \citenamefont {Jarillo-Herrero}, \citenamefont {Jacquod},\ and\ \citenamefont
  {LeRoy}}]{Yankowitz2012}%
  \BibitemOpen
  \bibfield  {author} {\bibinfo {author} {\bibfnamefont {M.}~\bibnamefont
  {Yankowitz}}, \bibinfo {author} {\bibfnamefont {J.~M.}\ \bibnamefont {Xue}},
  \bibinfo {author} {\bibfnamefont {D.}~\bibnamefont {Cormode}}, \bibinfo
  {author} {\bibfnamefont {J.~D.}\ \bibnamefont {Sanchez-Yamagishi}}, \bibinfo
  {author} {\bibfnamefont {K.}~\bibnamefont {Watanabe}}, \bibinfo {author}
  {\bibfnamefont {T.}~\bibnamefont {Taniguchi}}, \bibinfo {author}
  {\bibfnamefont {P.}~\bibnamefont {Jarillo-Herrero}}, \bibinfo {author}
  {\bibfnamefont {P.}~\bibnamefont {Jacquod}}, \ and\ \bibinfo {author}
  {\bibfnamefont {B.~J.}\ \bibnamefont {LeRoy}},\ }\href {\doibase
  10.1038/Nphys2272} {\bibfield  {journal} {\bibinfo  {journal} {Nature
  Physics}\ }\textbf {\bibinfo {volume} {8}},\ \bibinfo {pages} {382} (\bibinfo
  {year} {2012})}\BibitemShut {NoStop}%
\bibitem [{\citenamefont {Gorbachev}\ \emph {et~al.}(2014)\citenamefont
  {Gorbachev}, \citenamefont {Song}, \citenamefont {Yu}, \citenamefont
  {Kretinin}, \citenamefont {Withers}, \citenamefont {Cao}, \citenamefont
  {Mishchenko}, \citenamefont {Grigorieva}, \citenamefont {Novoselov},
  \citenamefont {Levitov},\ and\ \citenamefont {Geim}}]{Gorbachev2014}%
  \BibitemOpen
  \bibfield  {author} {\bibinfo {author} {\bibfnamefont {R.~V.}\ \bibnamefont
  {Gorbachev}}, \bibinfo {author} {\bibfnamefont {J.~C.}\ \bibnamefont {Song}},
  \bibinfo {author} {\bibfnamefont {G.~L.}\ \bibnamefont {Yu}}, \bibinfo
  {author} {\bibfnamefont {A.~V.}\ \bibnamefont {Kretinin}}, \bibinfo {author}
  {\bibfnamefont {F.}~\bibnamefont {Withers}}, \bibinfo {author} {\bibfnamefont
  {Y.}~\bibnamefont {Cao}}, \bibinfo {author} {\bibfnamefont {A.}~\bibnamefont
  {Mishchenko}}, \bibinfo {author} {\bibfnamefont {I.~V.}\ \bibnamefont
  {Grigorieva}}, \bibinfo {author} {\bibfnamefont {K.~S.}\ \bibnamefont
  {Novoselov}}, \bibinfo {author} {\bibfnamefont {L.~S.}\ \bibnamefont
  {Levitov}}, \ and\ \bibinfo {author} {\bibfnamefont {A.~K.}\ \bibnamefont
  {Geim}},\ }\href {\doibase 10.1126/science.1254966} {\bibfield  {journal}
  {\bibinfo  {journal} {Science}\ }\textbf {\bibinfo {volume} {346}},\ \bibinfo
  {pages} {448} (\bibinfo {year} {2014})},\ \Eprint
  {http://arxiv.org/abs/1409.0113} {1409.0113} \BibitemShut {NoStop}%
\bibitem [{\citenamefont {Yao}\ \emph {et~al.}(2008)\citenamefont {Yao},
  \citenamefont {Xiao},\ and\ \citenamefont {Niu}}]{Yao2008}%
  \BibitemOpen
  \bibfield  {author} {\bibinfo {author} {\bibfnamefont {W.}~\bibnamefont
  {Yao}}, \bibinfo {author} {\bibfnamefont {D.}~\bibnamefont {Xiao}}, \ and\
  \bibinfo {author} {\bibfnamefont {Q.}~\bibnamefont {Niu}},\ }\href {\doibase
  ARTN 235406 10.1103/PhysRevB.77.235406} {\bibfield  {journal} {\bibinfo
  {journal} {Physical Review B}\ }\textbf {\bibinfo {volume} {77}},\ \bibinfo
  {pages} {235406} (\bibinfo {year} {2008})}\BibitemShut {NoStop}%
\bibitem [{\citenamefont {Lensky}\ \emph {et~al.}(2015)\citenamefont {Lensky},
  \citenamefont {Song}, \citenamefont {Samutpraphoot},\ and\ \citenamefont
  {Levitov}}]{Lensky2015}%
  \BibitemOpen
  \bibfield  {author} {\bibinfo {author} {\bibfnamefont {Y.~D.}\ \bibnamefont
  {Lensky}}, \bibinfo {author} {\bibfnamefont {J.~C.}\ \bibnamefont {Song}},
  \bibinfo {author} {\bibfnamefont {P.}~\bibnamefont {Samutpraphoot}}, \ and\
  \bibinfo {author} {\bibfnamefont {L.~S.}\ \bibnamefont {Levitov}},\ }\href
  {\doibase 10.1103/PhysRevLett.114.256601} {\bibfield  {journal} {\bibinfo
  {journal} {Physical Review Letters}\ }\textbf {\bibinfo {volume} {114}},\
  \bibinfo {pages} {256601} (\bibinfo {year} {2015})},\ \Eprint
  {http://arxiv.org/abs/1412.1808} {1412.1808} \BibitemShut {NoStop}%
\bibitem [{\citenamefont {Semenoff}(1984)}]{Semenoff1984}%
  \BibitemOpen
  \bibfield  {author} {\bibinfo {author} {\bibfnamefont {G.~W.}\ \bibnamefont
  {Semenoff}},\ }\href {\doibase 10.1103/PhysRevLett.53.2449} {\bibfield
  {journal} {\bibinfo  {journal} {Physical Review Letters}\ }\textbf {\bibinfo
  {volume} {53}},\ \bibinfo {pages} {2449} (\bibinfo {year}
  {1984})}\BibitemShut {NoStop}%
\bibitem [{\citenamefont {Silva}\ \emph {et~al.}(2019)\citenamefont {Silva},
  \citenamefont {Jiménez-Galán}, \citenamefont {Amorim}, \citenamefont
  {Smirnova},\ and\ \citenamefont {Ivanov}}]{Silva2019}%
  \BibitemOpen
  \bibfield  {author} {\bibinfo {author} {\bibfnamefont {R.~E.~F.}\
  \bibnamefont {Silva}}, \bibinfo {author} {\bibfnamefont {A.}~\bibnamefont
  {Jiménez-Galán}}, \bibinfo {author} {\bibfnamefont {B.}~\bibnamefont
  {Amorim}}, \bibinfo {author} {\bibfnamefont {O.}~\bibnamefont {Smirnova}}, \
  and\ \bibinfo {author} {\bibfnamefont {M.}~\bibnamefont {Ivanov}},\ }\href
  {\doibase 10.1038/s41566-019-0516-1} {\bibfield  {journal} {\bibinfo
  {journal} {Nature Photonics}\ }\textbf {\bibinfo {volume} {13}},\ \bibinfo
  {pages} {849} (\bibinfo {year} {2019})}\BibitemShut {NoStop}%
\bibitem [{\citenamefont {Chacón}\ \emph {et~al.}(2020)\citenamefont
  {Chacón}, \citenamefont {Kim}, \citenamefont {Zhu}, \citenamefont {Kelly},
  \citenamefont {Dauphin}, \citenamefont {Pisanty}, \citenamefont {Maxwell},
  \citenamefont {Picón}, \citenamefont {Ciappina}, \citenamefont {Kim},
  \citenamefont {Ticknor}, \citenamefont {Saxena},\ and\ \citenamefont
  {Lewenstein}}]{Chacon2020}%
  \BibitemOpen
  \bibfield  {author} {\bibinfo {author} {\bibfnamefont {A.}~\bibnamefont
  {Chacón}}, \bibinfo {author} {\bibfnamefont {D.}~\bibnamefont {Kim}},
  \bibinfo {author} {\bibfnamefont {W.}~\bibnamefont {Zhu}}, \bibinfo {author}
  {\bibfnamefont {S.~P.}\ \bibnamefont {Kelly}}, \bibinfo {author}
  {\bibfnamefont {A.}~\bibnamefont {Dauphin}}, \bibinfo {author} {\bibfnamefont
  {E.}~\bibnamefont {Pisanty}}, \bibinfo {author} {\bibfnamefont {A.~S.}\
  \bibnamefont {Maxwell}}, \bibinfo {author} {\bibfnamefont {A.}~\bibnamefont
  {Picón}}, \bibinfo {author} {\bibfnamefont {M.~F.}\ \bibnamefont
  {Ciappina}}, \bibinfo {author} {\bibfnamefont {D.~E.}\ \bibnamefont {Kim}},
  \bibinfo {author} {\bibfnamefont {C.}~\bibnamefont {Ticknor}}, \bibinfo
  {author} {\bibfnamefont {A.}~\bibnamefont {Saxena}}, \ and\ \bibinfo {author}
  {\bibfnamefont {M.}~\bibnamefont {Lewenstein}},\ }\href {\doibase
  10.1103/PhysRevB.102.134115} {\bibfield  {journal} {\bibinfo  {journal}
  {Physical Review B}\ }\textbf {\bibinfo {volume} {102}},\ \bibinfo {pages}
  {134115} (\bibinfo {year} {2020})}\BibitemShut {NoStop}%
\bibitem [{\citenamefont {Yue}\ and\ \citenamefont
  {Gaarde}(2020)}]{Gaarde_PRL2020}%
  \BibitemOpen
  \bibfield  {author} {\bibinfo {author} {\bibfnamefont {L.}~\bibnamefont
  {Yue}}\ and\ \bibinfo {author} {\bibfnamefont {M.~B.}\ \bibnamefont
  {Gaarde}},\ }\href {\doibase 10.1103/PhysRevLett.124.153204} {\bibfield
  {journal} {\bibinfo  {journal} {Physical Review Letters}\ }\textbf {\bibinfo
  {volume} {124}},\ \bibinfo {pages} {153204} (\bibinfo {year}
  {2020})}\BibitemShut {NoStop}%
\bibitem [{\citenamefont {Moos}\ \emph {et~al.}(2020)\citenamefont {Moos},
  \citenamefont {Jürß},\ and\ \citenamefont {Bauer}}]{Bauer2020}%
  \BibitemOpen
  \bibfield  {author} {\bibinfo {author} {\bibfnamefont {D.}~\bibnamefont
  {Moos}}, \bibinfo {author} {\bibfnamefont {C.}~\bibnamefont {Jürß}}, \ and\
  \bibinfo {author} {\bibfnamefont {D.}~\bibnamefont {Bauer}},\ }\href@noop {}
  {\bibfield  {journal} {\bibinfo  {journal} {arXiv preprint arXiv:2007.13434}\
  } (\bibinfo {year} {2020})}\BibitemShut {NoStop}%
\bibitem [{\citenamefont {Wu}\ \emph {et~al.}(2018)\citenamefont {Wu},
  \citenamefont {Fatemi}, \citenamefont {Gibson}, \citenamefont {Watanabe},
  \citenamefont {Taniguchi}, \citenamefont {Cava},\ and\ \citenamefont
  {Jarillo-Herrero}}]{Herrero2018}%
  \BibitemOpen
  \bibfield  {author} {\bibinfo {author} {\bibfnamefont {S.}~\bibnamefont
  {Wu}}, \bibinfo {author} {\bibfnamefont {V.}~\bibnamefont {Fatemi}}, \bibinfo
  {author} {\bibfnamefont {Q.~D.}\ \bibnamefont {Gibson}}, \bibinfo {author}
  {\bibfnamefont {K.}~\bibnamefont {Watanabe}}, \bibinfo {author}
  {\bibfnamefont {T.}~\bibnamefont {Taniguchi}}, \bibinfo {author}
  {\bibfnamefont {R.~J.}\ \bibnamefont {Cava}}, \ and\ \bibinfo {author}
  {\bibfnamefont {P.}~\bibnamefont {Jarillo-Herrero}},\ }\href {\doibase
  10.1126/science.aan6003} {\bibfield  {journal} {\bibinfo  {journal}
  {Science}\ }\textbf {\bibinfo {volume} {359}},\ \bibinfo {pages} {76}
  (\bibinfo {year} {2018})}\BibitemShut {NoStop}%
\bibitem [{\citenamefont {Nagaosa}\ \emph {et~al.}(2010)\citenamefont
  {Nagaosa}, \citenamefont {Sinova}, \citenamefont {Onoda}, \citenamefont
  {MacDonald},\ and\ \citenamefont {Ong}}]{Nagaosa2010}%
  \BibitemOpen
  \bibfield  {author} {\bibinfo {author} {\bibfnamefont {N.}~\bibnamefont
  {Nagaosa}}, \bibinfo {author} {\bibfnamefont {J.}~\bibnamefont {Sinova}},
  \bibinfo {author} {\bibfnamefont {S.}~\bibnamefont {Onoda}}, \bibinfo
  {author} {\bibfnamefont {A.~H.}\ \bibnamefont {MacDonald}}, \ and\ \bibinfo
  {author} {\bibfnamefont {N.~P.}\ \bibnamefont {Ong}},\ }\href {\doibase
  10.1103/RevModPhys.82.1539} {\bibfield  {journal} {\bibinfo  {journal}
  {Reviews of Modern Physics}\ }\textbf {\bibinfo {volume} {82}},\ \bibinfo
  {pages} {1539} (\bibinfo {year} {2010})}\BibitemShut {NoStop}%
\bibitem [{\citenamefont {Sato}\ \emph {et~al.}(2019)\citenamefont {Sato},
  \citenamefont {Tang}, \citenamefont {Sentef}, \citenamefont {Giovannini},
  \citenamefont {Hübener},\ and\ \citenamefont {Rubio}}]{Rubio2019}%
  \BibitemOpen
  \bibfield  {author} {\bibinfo {author} {\bibfnamefont {S.~A.}\ \bibnamefont
  {Sato}}, \bibinfo {author} {\bibfnamefont {P.}~\bibnamefont {Tang}}, \bibinfo
  {author} {\bibfnamefont {M.~A.}\ \bibnamefont {Sentef}}, \bibinfo {author}
  {\bibfnamefont {U.~D.}\ \bibnamefont {Giovannini}}, \bibinfo {author}
  {\bibfnamefont {H.}~\bibnamefont {Hübener}}, \ and\ \bibinfo {author}
  {\bibfnamefont {A.}~\bibnamefont {Rubio}},\ }\href {\doibase
  10.1088/1367-2630/ab3acf} {\bibfield  {journal} {\bibinfo  {journal} {New
  Journal of Physics}\ }\textbf {\bibinfo {volume} {21}},\ \bibinfo {pages}
  {093005} (\bibinfo {year} {2019})}\BibitemShut {NoStop}%
\bibitem [{\citenamefont {McIver}\ \emph {et~al.}(2020)\citenamefont {McIver},
  \citenamefont {Schulte}, \citenamefont {Stein}, \citenamefont {Matsuyama},
  \citenamefont {Jotzu}, \citenamefont {Meier},\ and\ \citenamefont
  {Cavalleri}}]{Cavalleri2020}%
  \BibitemOpen
  \bibfield  {author} {\bibinfo {author} {\bibfnamefont {J.~W.}\ \bibnamefont
  {McIver}}, \bibinfo {author} {\bibfnamefont {B.}~\bibnamefont {Schulte}},
  \bibinfo {author} {\bibfnamefont {F.~U.}\ \bibnamefont {Stein}}, \bibinfo
  {author} {\bibfnamefont {T.}~\bibnamefont {Matsuyama}}, \bibinfo {author}
  {\bibfnamefont {G.}~\bibnamefont {Jotzu}}, \bibinfo {author} {\bibfnamefont
  {G.}~\bibnamefont {Meier}}, \ and\ \bibinfo {author} {\bibfnamefont
  {A.}~\bibnamefont {Cavalleri}},\ }\href {\doibase 10.1038/s41567-019-0698-y}
  {\bibfield  {journal} {\bibinfo  {journal} {Nat Phys}\ }\textbf {\bibinfo
  {volume} {16}},\ \bibinfo {pages} {38} (\bibinfo {year} {2020})}\BibitemShut
  {NoStop}%
\bibitem [{\citenamefont {Steinleitner}\ \emph {et~al.}(2017)\citenamefont
  {Steinleitner}, \citenamefont {Merkl}, \citenamefont {Nagler}, \citenamefont
  {Mornhinweg}, \citenamefont {Schüller}, \citenamefont {Korn}, \citenamefont
  {Chernikov},\ and\ \citenamefont {Huber}}]{Steinleitner2017}%
  \BibitemOpen
  \bibfield  {author} {\bibinfo {author} {\bibfnamefont {P.}~\bibnamefont
  {Steinleitner}}, \bibinfo {author} {\bibfnamefont {P.}~\bibnamefont {Merkl}},
  \bibinfo {author} {\bibfnamefont {P.}~\bibnamefont {Nagler}}, \bibinfo
  {author} {\bibfnamefont {J.}~\bibnamefont {Mornhinweg}}, \bibinfo {author}
  {\bibfnamefont {C.}~\bibnamefont {Schüller}}, \bibinfo {author}
  {\bibfnamefont {T.}~\bibnamefont {Korn}}, \bibinfo {author} {\bibfnamefont
  {A.}~\bibnamefont {Chernikov}}, \ and\ \bibinfo {author} {\bibfnamefont
  {R.}~\bibnamefont {Huber}},\ }\href {\doibase 10.1021/acs.nanolett.6b04422}
  {\bibfield  {journal} {\bibinfo  {journal} {Nano Letters}\ }\textbf {\bibinfo
  {volume} {17}},\ \bibinfo {pages} {1455} (\bibinfo {year}
  {2017})}\BibitemShut {NoStop}%
\bibitem [{\citenamefont {Wang}\ \emph {et~al.}(2018)\citenamefont {Wang},
  \citenamefont {Chernikov}, \citenamefont {Glazov}, \citenamefont {Heinz},
  \citenamefont {Marie}, \citenamefont {Amand},\ and\ \citenamefont
  {Urbaszek}}]{Bernhard2018}%
  \BibitemOpen
  \bibfield  {author} {\bibinfo {author} {\bibfnamefont {G.}~\bibnamefont
  {Wang}}, \bibinfo {author} {\bibfnamefont {A.}~\bibnamefont {Chernikov}},
  \bibinfo {author} {\bibfnamefont {M.~M.}\ \bibnamefont {Glazov}}, \bibinfo
  {author} {\bibfnamefont {T.~F.}\ \bibnamefont {Heinz}}, \bibinfo {author}
  {\bibfnamefont {X.}~\bibnamefont {Marie}}, \bibinfo {author} {\bibfnamefont
  {T.}~\bibnamefont {Amand}}, \ and\ \bibinfo {author} {\bibfnamefont
  {B.}~\bibnamefont {Urbaszek}},\ }\href {\doibase
  10.1103/RevModPhys.90.021001} {\bibfield  {journal} {\bibinfo  {journal}
  {Reviews of Modern Physics}\ }\textbf {\bibinfo {volume} {90}},\ \bibinfo
  {pages} {021001} (\bibinfo {year} {2018})}\BibitemShut {NoStop}%
\bibitem [{\citenamefont {Wang}\ \emph {et~al.}(2013)\citenamefont {Wang},
  \citenamefont {Steinberg}, \citenamefont {Jarillo-Herrero},\ and\
  \citenamefont {Gedik}}]{Wang2013}%
  \BibitemOpen
  \bibfield  {author} {\bibinfo {author} {\bibfnamefont {Y.~H.}\ \bibnamefont
  {Wang}}, \bibinfo {author} {\bibfnamefont {H.}~\bibnamefont {Steinberg}},
  \bibinfo {author} {\bibfnamefont {P.}~\bibnamefont {Jarillo-Herrero}}, \ and\
  \bibinfo {author} {\bibfnamefont {N.}~\bibnamefont {Gedik}},\ }\href
  {\doibase 10.1126/science.1239834} {\bibfield  {journal} {\bibinfo  {journal}
  {Science}\ }\textbf {\bibinfo {volume} {342}},\ \bibinfo {pages} {453}
  (\bibinfo {year} {2013})}\BibitemShut {NoStop}%
\bibitem [{\citenamefont {Higuchi}\ \emph {et~al.}(2017)\citenamefont
  {Higuchi}, \citenamefont {Heide}, \citenamefont {Ullmann}, \citenamefont
  {Weber},\ and\ \citenamefont {Hommelhoff}}]{Higuchi2017}%
  \BibitemOpen
  \bibfield  {author} {\bibinfo {author} {\bibfnamefont {T.}~\bibnamefont
  {Higuchi}}, \bibinfo {author} {\bibfnamefont {C.}~\bibnamefont {Heide}},
  \bibinfo {author} {\bibfnamefont {K.}~\bibnamefont {Ullmann}}, \bibinfo
  {author} {\bibfnamefont {H.~B.}\ \bibnamefont {Weber}}, \ and\ \bibinfo
  {author} {\bibfnamefont {P.}~\bibnamefont {Hommelhoff}},\ }\href {\doibase
  10.1038/nature23900} {\bibfield  {journal} {\bibinfo  {journal} {Nature}\
  }\textbf {\bibinfo {volume} {550}},\ \bibinfo {pages} {224} (\bibinfo {year}
  {2017})}\BibitemShut {NoStop}%
\bibitem [{\citenamefont {Houston}(1940)}]{Houston_1940}%
  \BibitemOpen
  \bibfield  {author} {\bibinfo {author} {\bibfnamefont {W.~V.}\ \bibnamefont
  {Houston}},\ }\href {\doibase 10.1103/PhysRev.57.184} {\bibfield  {journal}
  {\bibinfo  {journal} {Physical Review}\ }\textbf {\bibinfo {volume} {57}},\
  \bibinfo {pages} {184} (\bibinfo {year} {1940})}\BibitemShut {NoStop}%
\bibitem [{\citenamefont {Gierz}\ \emph {et~al.}(2013)\citenamefont {Gierz},
  \citenamefont {Petersen}, \citenamefont {Mitrano}, \citenamefont {Cacho},
  \citenamefont {Turcu}, \citenamefont {Springate}, \citenamefont
  {St{\"{o}}hr}, \citenamefont {K{\"{o}}hler}, \citenamefont {Starke},\ and\
  \citenamefont {Cavalleri}}]{Gierz2013_Snapshots}%
  \BibitemOpen
  \bibfield  {author} {\bibinfo {author} {\bibfnamefont {I.}~\bibnamefont
  {Gierz}}, \bibinfo {author} {\bibfnamefont {J.~C.}\ \bibnamefont {Petersen}},
  \bibinfo {author} {\bibfnamefont {M.}~\bibnamefont {Mitrano}}, \bibinfo
  {author} {\bibfnamefont {C.}~\bibnamefont {Cacho}}, \bibinfo {author}
  {\bibfnamefont {I.~C.}\ \bibnamefont {Turcu}}, \bibinfo {author}
  {\bibfnamefont {E.}~\bibnamefont {Springate}}, \bibinfo {author}
  {\bibfnamefont {A.}~\bibnamefont {St{\"{o}}hr}}, \bibinfo {author}
  {\bibfnamefont {A.}~\bibnamefont {K{\"{o}}hler}}, \bibinfo {author}
  {\bibfnamefont {U.}~\bibnamefont {Starke}}, \ and\ \bibinfo {author}
  {\bibfnamefont {A.}~\bibnamefont {Cavalleri}},\ }\href {\doibase
  10.1038/nmat3757} {\bibfield  {journal} {\bibinfo  {journal} {Nature
  Materials}\ }\textbf {\bibinfo {volume} {12}},\ \bibinfo {pages} {1119}
  (\bibinfo {year} {2013})},\ \Eprint {http://arxiv.org/abs/1304.1389}
  {1304.1389} \BibitemShut {NoStop}%
\bibitem [{\citenamefont {Li}\ \emph {et~al.}(2012)\citenamefont {Li},
  \citenamefont {Luo}, \citenamefont {Hupalo}, \citenamefont {Zhang},
  \citenamefont {Tringides}, \citenamefont {Schmalian},\ and\ \citenamefont
  {Wang}}]{Li2012}%
  \BibitemOpen
  \bibfield  {author} {\bibinfo {author} {\bibfnamefont {T.}~\bibnamefont
  {Li}}, \bibinfo {author} {\bibfnamefont {L.}~\bibnamefont {Luo}}, \bibinfo
  {author} {\bibfnamefont {M.}~\bibnamefont {Hupalo}}, \bibinfo {author}
  {\bibfnamefont {J.}~\bibnamefont {Zhang}}, \bibinfo {author} {\bibfnamefont
  {M.~C.}\ \bibnamefont {Tringides}}, \bibinfo {author} {\bibfnamefont
  {J.}~\bibnamefont {Schmalian}}, \ and\ \bibinfo {author} {\bibfnamefont
  {J.}~\bibnamefont {Wang}},\ }\href {\doibase 10.1103/PhysRevLett.108.167401}
  {\bibfield  {journal} {\bibinfo  {journal} {Phys Rev Lett}\ }\textbf
  {\bibinfo {volume} {108}},\ \bibinfo {pages} {167401} (\bibinfo {year}
  {2012})}\BibitemShut {NoStop}%
\bibitem [{\citenamefont {Huttner}\ \emph {et~al.}(2017)\citenamefont
  {Huttner}, \citenamefont {Kira},\ and\ \citenamefont {Koch}}]{Huttner2017}%
  \BibitemOpen
  \bibfield  {author} {\bibinfo {author} {\bibfnamefont {U.}~\bibnamefont
  {Huttner}}, \bibinfo {author} {\bibfnamefont {M.}~\bibnamefont {Kira}}, \
  and\ \bibinfo {author} {\bibfnamefont {S.~W.}\ \bibnamefont {Koch}},\ }\href
  {\doibase 10.1002/lpor.201700049} {\bibfield  {journal} {\bibinfo  {journal}
  {Laser {\&} Photonics Reviews}\ }\textbf {\bibinfo {volume} {11}},\ \bibinfo
  {pages} {1700049} (\bibinfo {year} {2017})}\BibitemShut {NoStop}%
\bibitem [{\citenamefont {Kelardeh}\ \emph {et~al.}(2017)\citenamefont
  {Kelardeh}, \citenamefont {Apalkov},\ and\ \citenamefont
  {Stockman}}]{Kelardeh2017_superlattice}%
  \BibitemOpen
  \bibfield  {author} {\bibinfo {author} {\bibfnamefont {H.~K.}\ \bibnamefont
  {Kelardeh}}, \bibinfo {author} {\bibfnamefont {V.}~\bibnamefont {Apalkov}}, \
  and\ \bibinfo {author} {\bibfnamefont {M.~I.}\ \bibnamefont {Stockman}},\
  }in\ \href {\doibase 10.1364/FIO.2017.JTu3A.21} {\emph {\bibinfo {booktitle}
  {Optics InfoBase Conference Papers}}},\ \bibinfo {series} {OSA Technical
  Digest (online)}, Vol.\ \bibinfo {volume} {Part F66-FiO 2017}\ (\bibinfo
  {publisher} {Optical Society of America},\ \bibinfo {year} {2017})\ p.\
  \bibinfo {pages} {JTu3A.21}\BibitemShut {NoStop}%
\bibitem [{\citenamefont {Kelardeh}\ \emph {et~al.}(2015)\citenamefont
  {Kelardeh}, \citenamefont {Apalkov},\ and\ \citenamefont
  {Stockman}}]{Kelardeh2015_Graphene}%
  \BibitemOpen
  \bibfield  {author} {\bibinfo {author} {\bibfnamefont {H.~K.}\ \bibnamefont
  {Kelardeh}}, \bibinfo {author} {\bibfnamefont {V.}~\bibnamefont {Apalkov}}, \
  and\ \bibinfo {author} {\bibfnamefont {M.~I.}\ \bibnamefont {Stockman}},\
  }\href {\doibase 10.1103/PhysRevB.91.045439} {\bibfield  {journal} {\bibinfo
  {journal} {Physical Review B - Condensed Matter and Materials Physics}\
  }\textbf {\bibinfo {volume} {91}},\ \bibinfo {pages} {45439} (\bibinfo {year}
  {2015})}\BibitemShut {NoStop}%
\bibitem [{\citenamefont {Stephanov}\ and\ \citenamefont
  {Yin}(2012)}]{Stephanov2012}%
  \BibitemOpen
  \bibfield  {author} {\bibinfo {author} {\bibfnamefont {M.~A.}\ \bibnamefont
  {Stephanov}}\ and\ \bibinfo {author} {\bibfnamefont {Y.}~\bibnamefont
  {Yin}},\ }\href {\doibase 10.1103/PhysRevLett.109.162001} {\bibfield
  {journal} {\bibinfo  {journal} {Physical Review Letters}\ }\textbf {\bibinfo
  {volume} {109}},\ \bibinfo {pages} {162001} (\bibinfo {year}
  {2012})}\BibitemShut {NoStop}%
\end{thebibliography}%

\end{document}